\documentclass[12pt,preprint]{aastex} 
 
\usepackage{natbib,epsfig}   

\newcommand{\hminhalf}{$h^{-1/2}$ }
\newcommand{\hminhalfnospace}{$h^{-1/2}$}
\newcommand{\etal}{et al. }
\newcommand{\degree}{{$^{\circ}$ }}
\newcommand{\degreenospace}{{$^{\circ}$}}
\newcommand{\msun} {M$_{\odot}$}
\newcommand{\rosat} {{\em ROSAT} }
\newcommand{\asca} {{\em ASCA} }
\newcommand{\bepposax} {{\em BeppoSAX} }
\newcommand{\XMM} {{\em XMM-Newton} }
\newcommand{\chandra} {{\em Chandra} }
\newcommand{\hunit} {{km s$^{-1}$ Mpc$^{-1}$} }

\begin{document} 
 
\title{An Unbiased Measurement of $H_0$ through Cosmic Background
  Imager Observations of the Sunyaev-Zel'dovich Effect in Nearby
  Galaxy Clusters}
\author{P.~S. Udomprasert, B.~S. Mason\altaffilmark{1}, A.~C.~S. Readhead, and
  T.~J. Pearson}
\affil{California Institute of Technology}
\affil{1200 East California Blvd, Pasadena, CA 91125}

\altaffiltext{1}{Current Address: National Radio Astronomy
  Observatory, P.O. Box 2, Green Bank, WV 24944}

\shorttitle{CBI observations of the SZE}
\shortauthors{Udomprasert \etal}
 
\begin{abstract}

We present $H_0$ results from Cosmic Background Imager (CBI)
observations of the Sunyaev-Zel'dovich Effect (SZE) in 7 galaxy
clusters, A85, A399, A401, A478, A754, A1651, and A2597.  These
observations are part of a program to study a complete, volume-limited
sample of low-redshift ($z<0.1$), X-ray selected clusters.  Our focus
on nearby objects allows us to study a well-defined, orientation
unbiased sample, minimizing systematic errors due to cluster
asphericity.
We use density models derived from \rosat imaging data and temperature
measurements from \asca and \bepposax spectral observations.  We
quantify in detail sources of error in our derivation of $H_0$,
including calibration of the CBI data, density and temperature models
from the X-ray data, Cosmic Microwave Background (CMB) primary
anisotropy fluctuations, and residuals from radio point source
subtraction.  From these 7 clusters we obtain a result of $H_0 =
67^{+30}_{-18}\mbox{}^{+15}_{-6}$ \hunit for an unweighted sample
average.
The respective quoted errors are random and systematic uncertainties
at 68\% confidence.  The dominant source of error is confusion from
intrinsic anisotropy fluctuations.

\end{abstract}

\section{Introduction}
\label{section:intro}

The Sunyaev-Zel'dovich Effect (SZE) is a distortion in the Cosmic
Microwave Background (CMB) spectrum caused by the scattering of CMB
photons by electrons in a hot gas such as that in galaxy
clusters\citep{sz2}.  When coupled with X-ray observations, the SZE
provides a direct measurement of $H_0$ which is independent of the
astronomical distance ladder.  The SZE is proportional to $\int n_eT_e
\, dl$, while the X-ray emission due to thermal bremsstrahlung is
proportional to $\int n_e^2\Lambda(E,T_e) \, dl$, where $n_e$ is the
electron density, $T_e$ is the electron temperature, and
$\Lambda(E,T_e)$ is the X-ray spectral emissivity, a function of the
energy of observation, $E$, and electron temperature; $dl$ indicates
integration along the line of sight through the cluster.  X-ray
imaging observations constrain the cluster density profiles, while
X-ray spectroscopy provides temperature measurements, allowing one to
predict the expected SZE towards a cluster.  The comparison of the
X-ray and SZE observations, coupled with the assumption that clusters
are spherically symmetric, yields $H_0$.  Improved radio observing
techniques \citep[e.g.,][]{MasonSZ, Viper3667, MITO, RyleH0,
Reese2002} are providing increasingly precise measurements of the SZE,
while new X-ray data from \chandra and \XMM
\citep[e.g.,][]{a2029_chandra2,a478_chandra,xmm_cluster,xmm_a1835} are
yielding spatially resolved temperature measurements.  Together, these
observations will allow more accurate direct measurements of $H_0$
from combined SZE and X-ray data.

The most serious systematic sources of error in deriving $H_0$ from a
joint SZE and X-ray analysis are deviations from the assumptions that
cluster gas is spherical, smooth, and isothermal.  Several studies
\citep[e.g.,][]{cm1980,einstein_morph1,einstein_morph2} show that many
clusters are aspherical, while X-ray data demonstrate that temperature
profiles may not be isothermal \citep{MFSV98,DM2002}, nor is the gas
smooth, particularly in the case of clusters that have recently
merged.  One can best deal with these difficulties by observing nearby
clusters (as has been done by \citealt{MyersSZ, MasonSZ, Viper3667,
MITO}), where a complete sample of clusters can be defined with
confidence.
By averaging the $H_0$ results over all clusters in an
orientation-unbiased sample, any systematic errors due to cluster
asphericity can be minimized.  Also, the larger angular sizes of
nearby clusters allow one to study more easily the effects of a
non-smooth and non-isothermal gas.

This paper reports first results from a program to measure the SZE in
a complete sample of low-redshift ($z<0.1$) galaxy clusters using the
Cosmic Background Imager (CBI), a 13-element interferometer located in
the Chilean Andes.
The CBI is ideal for observing low-$z$ clusters with high resolution
and sensitivity, and we take advantage of these capabilities to
overcome some of the difficulties described above.  In this paper, we
report a preliminary value of $H_0$ from 7 clusters based on CBI
observations and published X-ray data from \rosat \citep{MM2000},
\asca \citep{MFSV98,White2000}, and \bepposax \citep{DM2002}.  Results
including temperature profiles from \chandra and \XMM
observations will be presented in a future paper.

We first describe the CBI observations and calibration, and the
analysis used to determine $H_0$ from our SZE observations and
published X-ray data.  We discuss sources of error, including
observational errors from calibration accuracy, thermal noise, primary
anisotropy fluctuations in the CMB, and residuals from point source
subtraction.  We also quantify errors from model-dependent sources
such as cluster density profiles and electron temperature.  Finally we
discuss possible errors from the assumptions that the cluster gas has
a smooth and isothermal distribution, and we consider ways to improve
upon these results.  Throughout the paper, we use $H_0 = 100 h$ km
s$^{-1}$ Mpc$^{-1}$, and we assume a flat $\Lambda$-CDM universe with
$\Omega_m$=0.3, $\Omega_{\Lambda}$=0.7.

\section{Observations}

The CBI is a 13-element radio interferometer operating in ten 1 GHz
frequency channels from 26 GHz to 36 GHz.  Dishes 0.9 m in diameter
are mounted on a 6 m tracking platform, allowing a range of baselines
from 1 m to 5.5 m.  The instantaneous field of view of the instrument
is $45'$ FWHM and its resolution ranges from $3'$ to $10'$, depending
on configuration.  The telescope has an altitude-azimuth mount, and
the antenna platform can also be rotated about the optical axis to
increase the aperture-plane ($u, v$) coverage.  The high
electron-mobility transistor (HEMT) amplifier receivers have noise
temperatures of $\sim 25$ K, and the typical system noise temperature,
including ground spillover and atmosphere, is $\sim 30$ K averaged
over all 10 bands.  The frequency of operation of the CBI was chosen
as a compromise between the effects of astronomical foregrounds,
atmospheric emission, and the sensitivity that can be achieved with
HEMT amplifiers.  Details of the instrument design may be found in
\citet{CBIPaper1,CBIInstr} and on the CBI web
site\footnote{http://www.astro.caltech.edu/$\sim$tjp/CBI/index.html}.

Nearby galaxy clusters have angular sizes of several arcminutes and
are ``resolved out'' by large interferometers with high resolution,
but they are ideal targets for the CBI, which has been optimized to
study CMB fluctuations on these angular scales.  At low redshift
($z<0.1$) it is feasible to define a complete, orientation unbiased
sample from X-ray data, so we take advantage of the exceptional
sensitivity of the CBI to nearby clusters to study a large,
volume-limited sample, minimizing bias from cluster asphericity.  To
compile our sample, we combined the results of \rosat cluster surveys
by \citet{XBACS,BCS}, \citet{REFLEX1}, and \citet{REFLEX_final}.  The
flux limit of our sample is $f_{0.1-2.4 {\rm keV}} > 1.0 \times
10^{-11}$ erg cm$^{-2}$ s$^{-1}$, which is significantly higher than
the expected completeness levels of these catalogs.  We then imposed a
redshift limit $z<0.1$, and selected a volume complete sample by only
including clusters with $L_{0.1-2.4 {\rm keV}} > 1.13 \times 10^{44}
h^{-2}$ erg s$^{-1}$.\footnote{Note that the X-ray surveys use
$h=0.5$.  We have converted their listed luminosities to $h=1.0$.}
Due to the telescope elevation limit of $>43$\degree and latitude of
$-$23\degreenospace, we are restricted to observing sources with
declinations $-$70\degreenospace $<\delta <24$\degreenospace.  Our
sample contains 24 clusters that are observable with the CBI.  These
are listed in Table \ref{tbl:sample}.  We have noted in the table
which clusters have public \rosat and \asca data available, as well as
which clusters have been or are scheduled to be targeted with the \XMM
and \chandra observatories.  The 15 most luminous clusters still
constitute a volume-complete sample, and since they have near complete
X-ray observations available, we define this group to be our primary
sample.  In this paper, we focus on 7 of these clusters, A85, A399,
A401, A478, A754, A1651, and A2597, which represent a range of X-ray
luminosities in the sample.

The CBI has been fully operational since January 2000, and the
clusters presented here were observed during the period from January
2000 to May 2001.  The CBI 1 meter baselines are most sensitive to
emission on $\sim$ 30$'$ scales, so contributions from the sun and
moon are potential contaminants.  To avoid the sun, we observe only at
night, and observations of CMB and SZE fields are used only if the
angular separation from the moon is at least 60$^{\circ}$.  We
estimate any residual contamination from the moon to be $< 1-2 \mu$K.

Our observing strategy has been designed to remove contamination from
ground spillover, which on 1 meter baselines generally contributes
between a few tens and a few hundreds of mJy of signal, but can be as
high as a few Jy.  Spillover is most severe for sources at low
elevation when the fringe pattern on some of the short baselines can
remain roughly parallel to the horizon as we track a source.  The
ground signal remains constant on hour timescales, and a differencing
scheme accurately removes the contamination.  For observations of CMB
fields, the CBI employs a Lead-Trail (L-T) differencing scheme where 2
fields are observed in succession at the same hour angle and
subtracted from each other.  This removes the ground signal and any
other potential spurious signals, with the level of potential
residuals being $<1.3\%$ of the primary anisotropy signal
\citep{CBIInstr}.  However, the differencing increases the noise by a
factor of $\sqrt{2}$.  The CMB primary anisotropies are the largest
source of contamination in the SZE observations, and L-T differencing
would also increase this source of noise by $\sqrt{2}$.  For cluster
observations, we therefore use a Lead-Main-Trail (LMT) differencing
scheme where an average of the lead and trail fields is subtracted
from the main field.  This increases the required observing time by
50\%, but it reduces the increase in CMB noise from $\sqrt{2}$ to
$\sqrt{3/2}$.  While we have not measured the level of residual
contamination from the LMT differencing scheme, it should be better
than that from L-T differencing, since any linear changes in time will
cancel out in the Lead and Trail fields.  Furthermore, any residual
signal will be much smaller than the primary anisotropy signal, and
would be negligible by comparison.
We used the NRAO VLA Sky Survey \citep{NVSS} to select Lead and Trail
fields where the contamination from point sources was minimized.  LMT
separations range from 9 minutes to 16.5 minutes in right ascension.
Table \ref{tbl:pos} lists the Lead, Main, and Trail pointing positions
used for each cluster, as well as total observation times (L+M+T
combined) and rms noise levels in the maps.  The positions of the main
field were taken from \citet{XBACS}.  These positions are obtained
from \rosat All-Sky Survey data, and they often differ from centroid
derived from pointed PSPC and HRI observations by $\sim 1'$, a
substantial fraction of the CBI 5$'$ synthesized beam.  Where
available, we use the centroid positions from the pointed observations
in our analysis.

\subsection{Calibration and Data Editing}
\label{section:data}

The data were calibrated through nightly observations of one or more
primary flux calibrators, Taurus A, Virgo A, Jupiter, or Saturn, which
were chosen for their brightness and lack of variability at the CBI 
frequencies.  A set of secondary flux calibrators (3C279, 3C273,
J1743$-$038, B1830$-$210, and J1924$-$292) were observed regularly and were
used to calibrate the data on nights when all primary calibrators were
not visible or were too close to the moon to be observed.
On those nights, the flux densities of the secondary calibrators were
bootstrapped from observations of the primary calibrators which were
nearest in time.  The CBI flux density scale is based on single dish
measurements of Jupiter, showing $T_{\rm Jup} = 152 \pm 5$ K at 32 GHz
\citep{MasonJup}.  Jupiter has a non-thermal spectrum across the CBI
bandpass, so we determined flux scales for the other 9 channels by
transferring the Jupiter 32 GHz flux to TauA, which has a known
power-law spectrum of $\alpha=-0.299$ \citep{Baars77}, where
$S_{\nu}\propto \nu^{\alpha}$.  We estimate that there is a 5\%
systematic uncertainty in our absolute calibration
\citep{CBIPaper1,CBIPaper2}.

We bracketed each of the LMT cluster scans and calibrator scans with
measurements of an internal noise source, whose equivalent flux
density at each baseline and channel is referenced from the celestial
primary calibrators.  We originally intended to use the noise source
to remove instrumental gain fluctuations throughout the night, but we
found that the gain fluctuations ($\sim$ 3\% rms variations) are more
stable than the fluctuations in the noise source itself, so all of the
noise source measurements were averaged together over the night.  The
individual baselines were then rescaled to give the same response.
This removed baseline-based gain and phase calibration errors, but
introduced antenna based amplitude errors which were removed through
the subsequent primary flux density calibration.  At the beginning and
end of each night we also performed a quadrature calibration to
measure the gain and phase offsets between the real and imaginary
outputs from the correlator.  The rms quadrature phase is $\sim$
5\degree, and the rms gain error is $\sim 10\%$.  The corrections were
stable over timescales of several weeks.  Observations of the primary
calibrators during the year 2000 showed random errors in calibration
of 3\% night-to-night.  All the clusters were observed over at least 5
nights, so the maximum expected random calibration error for each
cluster is 3\%$/\sqrt{5} = 1.3$\%.

To improve our aperture $(u,v)$ coverage, we rotate the deck about the
telescope optical axis by 10\degree between LMT groups.  By changing
the orientation of the telescope with respect to the source, we were
able to reduce any false signals which were generated in the receiver
electronics.  We also observed a bright ($\sim$ 1 Jy) source near the
fields ($\lesssim20^{\circ}$) at each deck angle, which allowed us to
determine the magnitude of pointing errors.  We found that the
absolute rms radio pointing was $\sim$ 22$''$, and the rms tracking
errors were $\sim$2$''$.  These errors are very small compared to the
CBI 45$'$ primary beam and the 4$'$ synthesized beam, and we have
performed Monte Carlo simulations which show that random pointing
errors of this magnitude do not bias our $H_0$ results.

Data editing was done both automatically and manually.  The telescope
control system automatically flagged data taken during periods when
the data may have been unreliable, such as when the telescope was not
tracking properly, a receiver was warm, a local oscillator was not
phase locked, or the total power of a receiver was outside the normal
range.  Notes in the observer log were used to examine periods where
there were instrumental problems or bad weather, indicated by visible
cloud cover or corrupted visibilities on the short baselines.  Two
percent of the data were removed on the basis of these inspections.
Occasionally, we still saw signals from instrumental glitches or from
the atmosphere during less optimal weather.  To reject these
observations, we filtered out data with amplitudes that differed from
the scan mean by more than five times the scan rms.  This criterion
rejected a negligible fraction of the data, and our results are not
sensitive to the precise level of the cut.  We also filtered out data
whose scatter was more than two times the noise expected based on the
integration time and the system properties.  This rejected less than
0.1\% of the data.

\begin{figure*}
\plotone{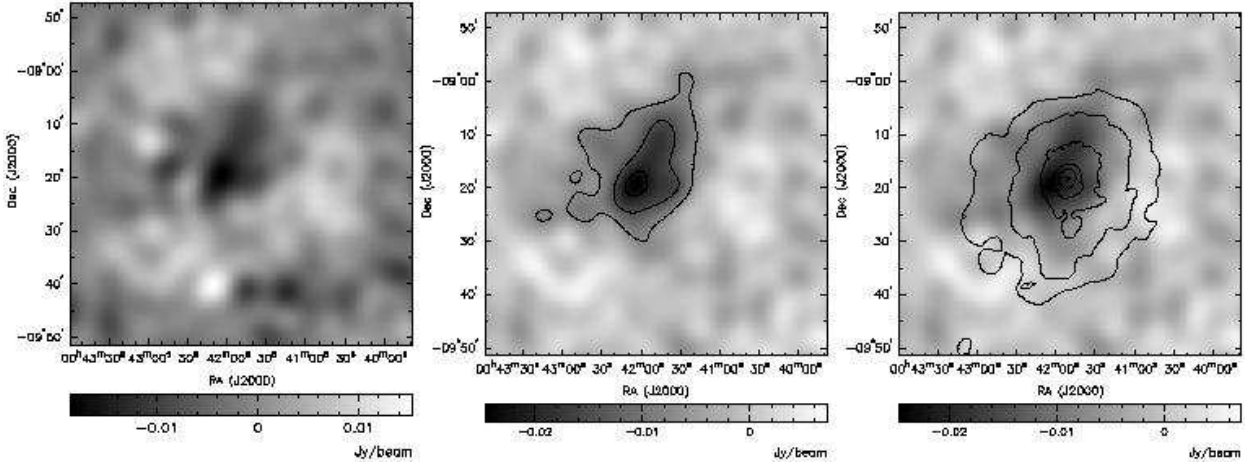}
\caption{The figure on the left is a dirty CBI image of A85 before
  point source subtraction.  The center figure shows the point sources
  subtracted, and the image has been convolved with a 5$'$ Gaussian
  restoring beam.  The SZE contour levels in the center plot are
  -0.0074, -0.015, -0.022 Jy beam$^{-1}$ (30\%, 60\%, 90\% of the peak
  of -0.0246 Jy beam$^{-1}$).  The figure on the right shows the same
  grayscale image with \rosat PSPC contours overlaid.  The X-ray
  contour levels in the plot are 0.0005, 0.001, 0.005, 0.02, 0.05,
  0.35, 0.5 counts s$^{-1}$ pixel $^{-1}$.  }
\label{fig:a85_map}
\end{figure*}

\begin{figure*}
\plotone{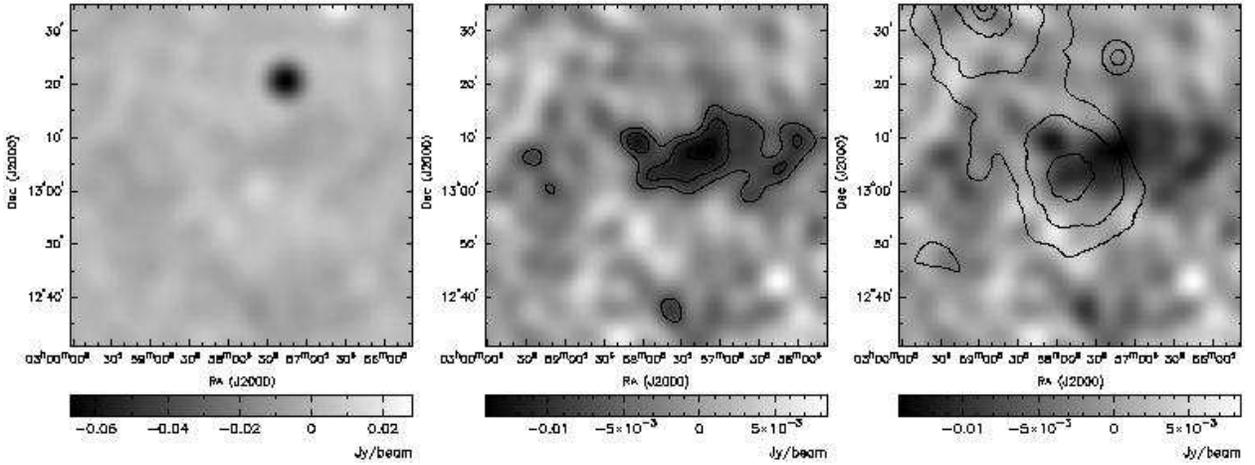}
\caption{A399: Same as in Figure \ref{fig:a85_map}.  The X-ray
  contours show A399's companion, A401, which does not appear in the
  SZE map due to attenuation by the CBI primary beam. The SZE contour
  levels in the center plot are -0.007, -0.010, -0.013 Jy beam$^{-1}$
  (50\%, 70\%, 90\% of the peak of -0.0141 Jy beam$^{-1}$).  The X-ray
  contour levels in the plot are 0.0005, 0.001, 0.005, 0.02, 0.05
  counts s$^{-1}$ pixel $^{-1}$.   }
\label{fig:a399_map}
\end{figure*}

\begin{figure*}
\plotone{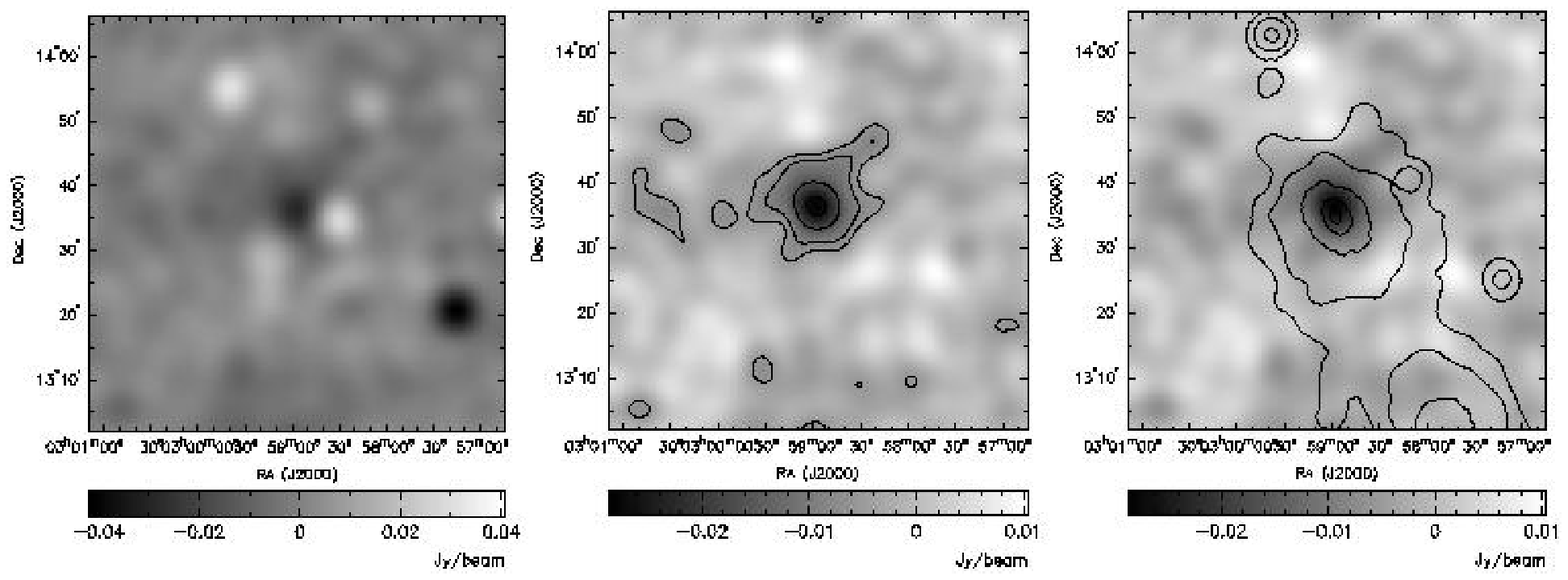}
\caption{A401: Same as in Figure \ref{fig:a399_map}.  The SZE contour
  levels in the center plot are -0.0055, -0.0086, -0.017, -0.026 Jy
  beam$^{-1}$ (20\%, 30\%, 60\%, 90\% of the peak of -0.0287 Jy
  beam$^{-1}$).  The X-ray contour levels in the plot are 0.0005,
  0.001, 0.005, 0.02, 0.05 counts s$^{-1}$ pixel $^{-1}$.  }
\label{fig:a401_map}
\end{figure*}

\begin{figure*}
\plotone{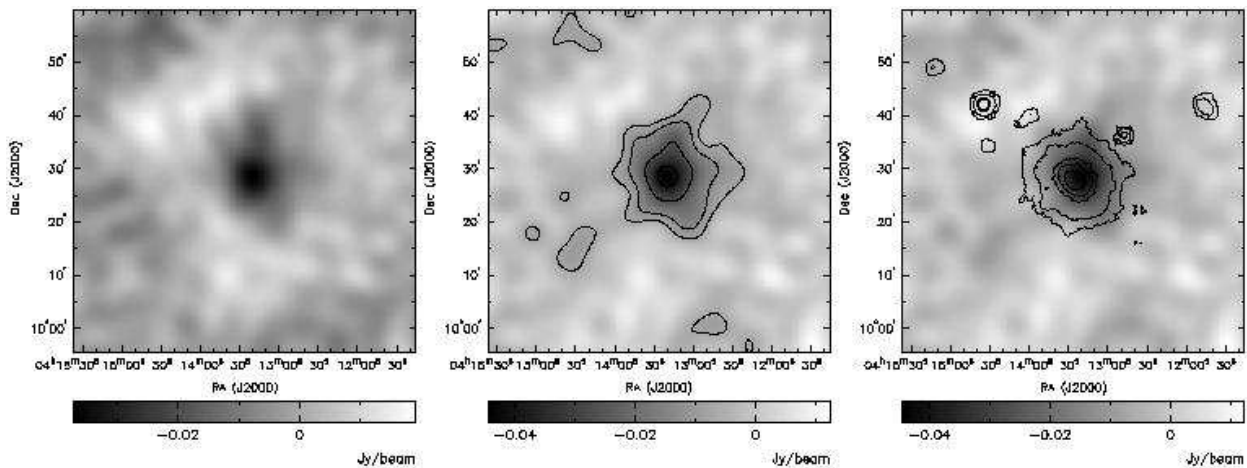}
\caption{A478: Same as in Figure \ref{fig:a85_map}.  The SZE contour
  levels in the center plot are -0.007, -0.013, -0.0266, -0.04 Jy beam$^{-1}$
  (15\%, 30\%, 60\%, 90\% of the peak of -0.0444 Jy beam$^{-1}$).  The X-ray
  contour levels in the plot are 0.0005, 0.001, 0.005, 0.01, 0.03
  counts s$^{-1}$ pixel $^{-1}$.  }
\label{fig:a478_map}
\end{figure*}

\begin{figure*}
\plotone{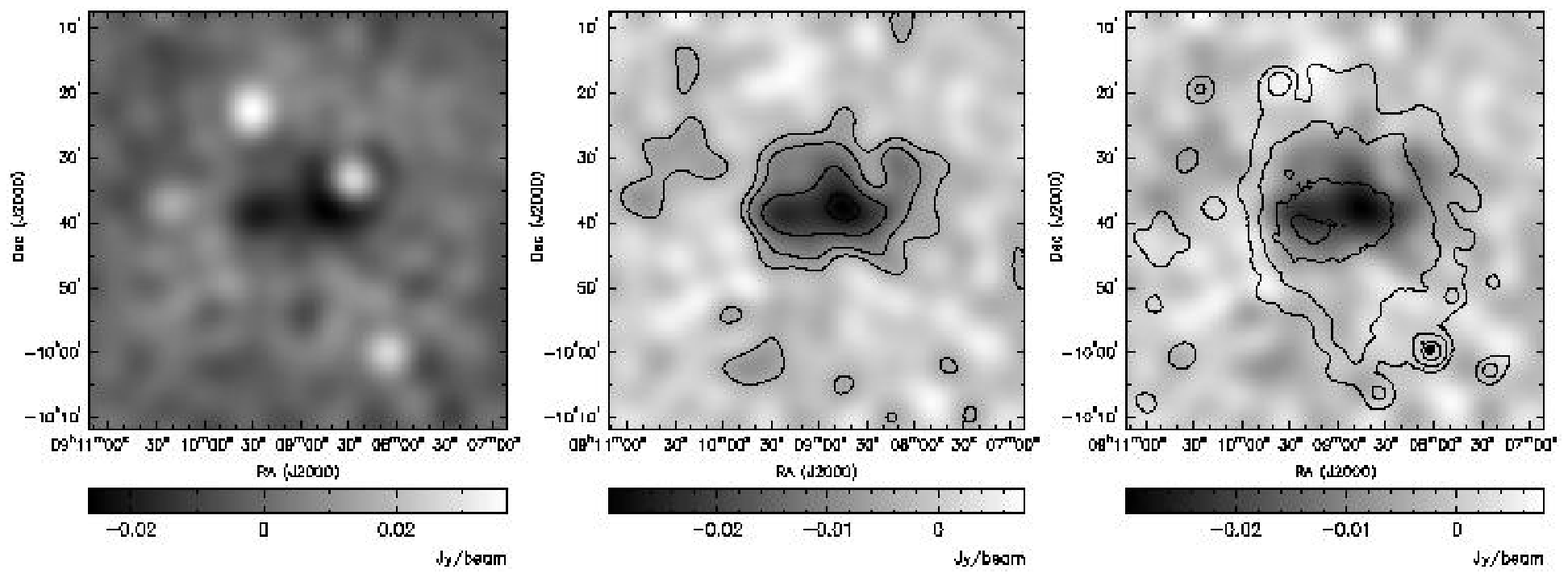}
\caption{A754: Same as in Figure \ref{fig:a85_map}.  The SZE contour
  levels in the center plot are -0.0045, -0.009, -0.0178, -0.0267 Jy beam$^{-1}$
  (15\%, 30\%, 60\%, 90\% of the peak of -0.0297 Jy beam$^{-1}$).  The X-ray
  contour levels in the plot are 0.0005, 0.001, 0.005, 0.02, 0.05, 0.2
  counts s$^{-1}$ pixel $^{-1}$. }
\label{fig:a754_map}
\end{figure*}

\begin{figure*}
\plotone{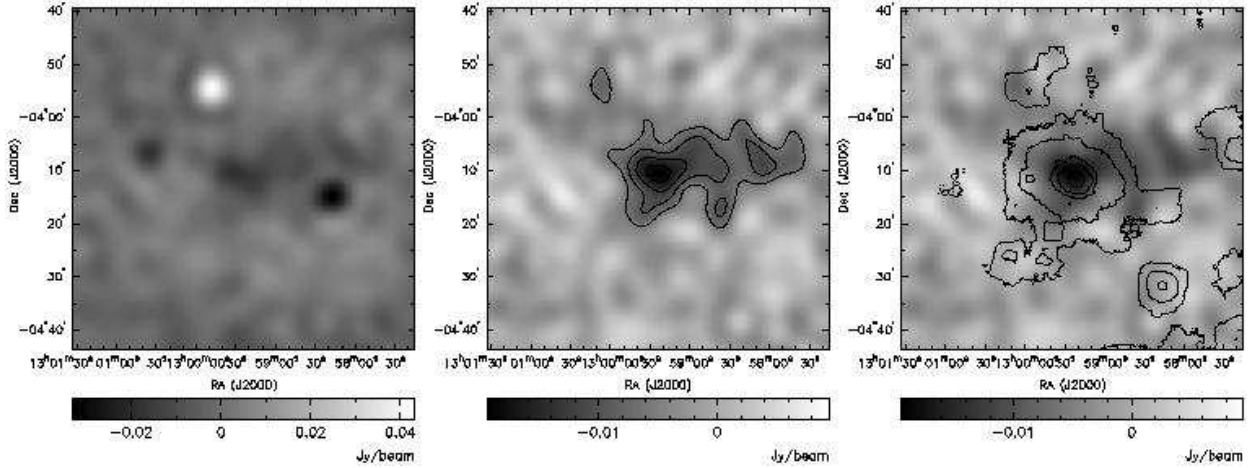}
\caption{A1651: Same as in Figure \ref{fig:a85_map}.  The SZE contour
  levels in the center plot are -0.006, -0.0097, -0.0136, -0.0175 Jy beam$^{-1}$
  (30\%, 50\%, 70\%, 90\% of the peak of -0.0194 Jy beam$^{-1}$).  The X-ray
  contour levels in the plot are 0.0005, 0.001, 0.005, 0.02, 0.05, 0.3
  counts s$^{-1}$ pixel $^{-1}$.}
\label{fig:a1651_map}
\end{figure*}

\begin{figure*}
\plotone{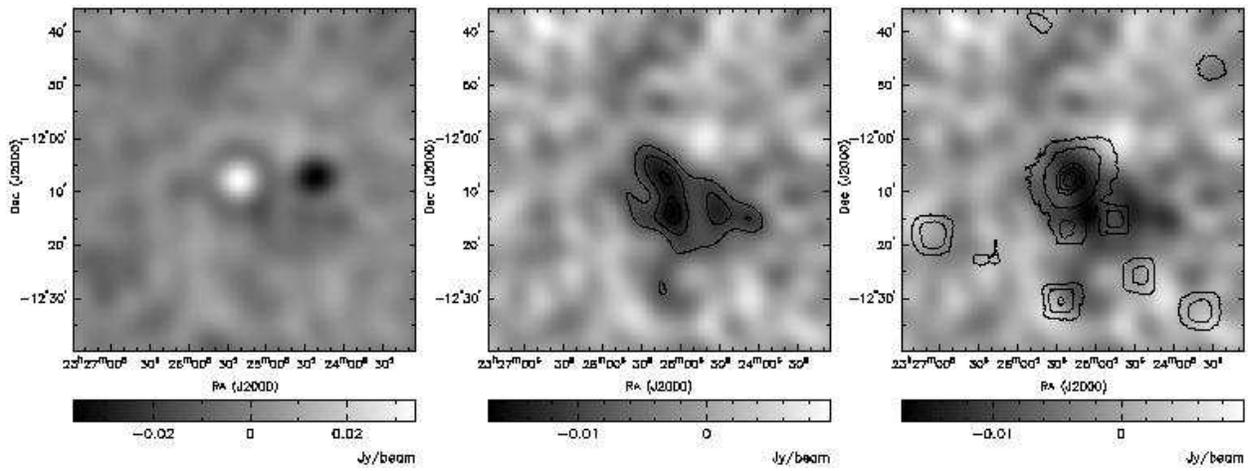}
\caption{A2597: Same as in Figure \ref{fig:a85_map}.  The SZE contour
  levels in the center plot are -0.0085 -0.0119 -0.0153 Jy beam$^{-1}$
  (50\%, 70\%, 90\% of the peak of -0.017 Jy beam$^{-1}$).  The X-ray
  contour levels in the plot are 0.0005, 0.001, 0.005, 0.02, 0.05, 0.3
  counts s$^{-1}$ pixel $^{-1}$.}
\label{fig:a2597_map}
\end{figure*}

Radio point sources present in all of the observations were subtracted
through a combination of fitting on the CBI long baselines ($>2.5$ m)
and observations with the OVRO 40-m telescope.  We used the 40-m
telescope to measure the fluxes of all NVSS sources within 10$'$ of
the LMT pointing centers.  In each cluster there were between 10 to 21
such sources, of which we detected between 1--5 per cluster whose flux
had to be subtracted from the CBI visibility data.  Outside a 10$'$
radius from the pointing centers, we fit for the fluxes of known NVSS
sources using the long CBI baselines.  Between 7--16 sources were
detected in each cluster at the 2.5-$\sigma$ level. Details of the
point source subtraction are presented in Section \ref{section:ptsrc}.
Figures \ref{fig:a85_map} to \ref{fig:a2597_map} show A85, A399, A401,
A478, A754, A1651, and A2597 before and after point source
subtraction.  The left hand figures are dirty images, and point
sources have not been subtracted.  The dirty images are often
dominated by a small number of bright sources that can mask the
presence of both the cluster and fainter point sources.  The middle
figures show the clusters after point source subtraction, and the
images have been deconvolved.  The right hand figures show the
grayscale from the deconvolved maps with X-ray contours from \rosat
PSPC data \citep{MM2000}.

\subsection{Notes on Individual Clusters}

\subsubsection{A85}

We observed A85 over 11 nights between July and December 2000.  X-ray
observations show that A85 has a central cooling flow, but this
cluster also shows signs of merger activity.  There is a smaller group
of galaxies just south of the cluster, which can be seen in the X-ray
contours.  \asca and \bepposax temperature maps show that this
``southern blob'' is slightly hotter than the rest of the cluster,
indicating that it is likely interacting with the cluster, rather than
a foreground projection \citep{MFSV98,a85_bepposax}.  If the
subcluster were independent, one would expect it to be cooler than the
main cluster given its smaller size.  \citet{a85_chandra} study the
merger of the subcluster in detail through \chandra observations.
\citet{a85_bepposax} determine an overall temperature for the cluster
of 6.6$\pm$0.3 keV, in agreement with the \asca and \citet{DM2002}
results.

VLA observations show some extended emission from a very steep
spectrum radio source (VSSRS) just southwest of the cluster center at
333 MHz \citep{a85_halo,a85_bepposax}.  Although there is a brighter
patch in the CBI map at this location, one would not expect to see
emission from the VSSRS at 31 GHz.  \citet{a85_halo} measure a flux of
3.15$\pm$0.15 Jy at 326.5 MHz.  If we extrapolate the spectral index
of $\alpha = -2.97$ between 300 MHz and 3 GHz, we expect a flux at 31
GHz of 4 $\mu$Jy, and the bright blob in the CBI map is more likely a
CMB hot spot.  The presence of the VSSRS, however, indicates a
possibility of Compton scattering from relativistic non-thermal
electrons in this region.  See Section \ref{section:nt} for details on
how this could affect our $H_0$ derivation.

We take the \rosat HRI centroid of 00:41:50.94,$-$9:18:10.7 (J2000)
\citep{a85_hri_cf} to be the location of SZE centroid in our fits.  Of
all the clusters presented here, A85 has the largest number of radio point
sources (21 at 1.4 GHz to 2.5 mJy) within 10$'$ of the cluster center.

\subsubsection{A399/A401}
We observed A399 and A401 over 6 nights during October and November 2000.
A399 and A401 are a pair of clusters which are close together on the
sky and in redshift.  X-ray observations indicate that the clusters
likely have interacted in the past or are currently interacting
\citep{a399_a401_asca,a399_a401_hri}.  The scenario favored by
\citet{a399_a401_hri} is that the clusters collided some time in the
past, disrupting their respective cooling flows, features which are
normally associated with clusters containing cD galaxies.  The
collision could also be responsible for the radio halo associated
with A401.  The halo has a steep spectrum $\alpha\sim -1.4$ and a total
flux density of 21 mJy at 1.4 GHz \citep{new_halo}.  Extrapolating the
spectral index to 31 GHz, the halo would have a flux of 0.3 mJy at the
CBI frequencies and is not expected to be a significant source of
contamination.  A nonthermal SZE from the halo electrons is possible
but difficult to quantify (see Section \ref{section:nt}).

The clusters are separated by only about 30$'$, which is smaller than
the CBI primary beam.  The primary beam attenuates the cluster signal
so much that the companions do not appear in the respective maps, but
we take into account the presence of the A401 when fitting for $H_0$
from the A399 data, and vice versa.

The cluster pair has several very bright radio sources in the field of
view and appear very ``dirty'' in the unsubtracted CBI maps.  However,
those sources can be accurately fitted out.  (See Section
\ref{section:ptsrc} for details).  There is a bright spot SW of A399
that appears in the A399 map at 10.1 mJy which does not correspond to
any NVSS sources.  It is possible this is an inverted spectrum source
which falls below the NVSS detection limit at 1.4 GHz.  If that is the
case, its spectral index would be $\alpha > 0.45$, which is reasonable
considering the distribution of spectral indices we discuss in Section
\ref{section:ptsrc}.  If we assume it is a genuine source and we fit
for its flux, our $H_0$ result changes by $<6$\%, a small amount
compared to the uncertainty from the CMB.

\subsubsection{A478}

We observed A478 over 11 nights during February, November, and
December 2000.  A478 is one of the most X-ray luminous clusters in the
sample.  It has very little point source contamination, and its X-ray
profile is extremely regular.  A478 is the largest cooling flow
cluster within $z<0.1$ \citep{a478_hri}, an indication of its relaxed
state.  From the \rosat HRI observations, \citet{a478_hri} find the
centroid of the X-ray emission to be RA=04:13:25.5, Dec=10:27:58
(J2000).  Although we have used slightly different coordinates as our
pointing center, in our analysis, we take this position to be the
centroid of the SZE emission as well.  A478 is one of the cleanest
clusters in our sample in terms of point source contamination.  There
are very few central sources, and a very small number of sources at
large radius whose fluxes need to be fitted.

\subsubsection{A754}

We observed A754 over 15 nights during February, October, and November
2000.  A754 is an irregular cluster which is considered to be a
prototypical merging system
\citep[e.g.,][]{a754_temp_map,a754_merger}.  Although we have included
this cluster in our sample to maintain completeness, we recognize that
such a disturbed cluster could contribute biases of its own.  In the
final sample, we therefore present values of $H_0$ with and without
A754.

The cluster is also known to have a strong radio halo.  At 74 MHz and
330 MHz, the halo flux is 4 Jy and 750 mJy, respectively
\citep{a754_vla_halo}.  \citet{new_halo} find a flux of 86 mJy at 1.4
GHz, and a spectral index of $\alpha=1.5$ from comparisons to
observations at 330 MHz.  Assuming this spectral index, we would
expect the halo flux at 31 GHz to be 0.8 mJy.  

A754 is very heavily contaminated with bright point sources.  One
source in particular
about
15$'$ NW of the cluster appears in the CBI map not to be well
subtracted.  This is most likely due to the fact that a spherical
$\beta$-model is not a good approximation to this elliptical disturbed
cluster, leaving a 10 mJy residual in the point source fit.
The level of subtraction for this particular source
does not change the value of $H_0$ determined for this cluster.

\subsubsection{A1651}

We observed A1651 over 8 nights during February 2000 and April and May
2001.  A1651 appears to be a dynamically relaxed cD cluster with a
regular \rosat PSPC profile \citep[e.g]{a1651_optical, MFSV98},
However, in their analysis of \asca observations, \citet{MFSV98} find
that a cooling component is not required in the fit.  This cluster
appears to be unremarkable, although it has a large number of bright
point sources which need to be subtracted from the CBI data.

\subsubsection{A2597}

We observed A2597 over 5 nights during September, October, and
November 2000.  A2597 is another regular cD cluster with a cooling
flow.  Its X-ray luminosity is among the weakest in our sample, and
the SZE maps indicate this.  Our detection is marginal at best, and
the error in deriving $H_0$ from this cluster is large.
\citet{a2597_rosat} take the centroid of the cluster to be the
position of the central cD galaxy (23:25:19.64,$-$12:07:27.4, J2000),
which we also use as the centroid in the SZE fits.  The cD galaxy is a
strong emitter at 31 GHz.  We determine a flux for the central source
of 40 mJy, both from the OVRO 40-m and the CBI long baselines ($>$ 3
m), giving us confidence in our measurement.

\section{Analysis Method}

\subsection{Modeling the Cluster Gas}
We assume that the cluster gas is well fitted by a spherical
isothermal $\beta$-model \citep{CF76} .  We will discuss possible
implications of this assumption later in the paper.  The gas distribution
is assumed to follow the form
\begin{equation}
n_e(r) = n_{e0}\left(1+\frac{r^2}{r_0^2}\right)^{-3\beta/2},
\label{eq:density}
\end{equation} 
where $n_{e0}$ is the central electron density, $r_0$ is the physical
core radius (related to the angular core radius, $\theta_0$, by
$r_0=D_A \theta_0,$ where $D_A$ is the angular diameter distance to
the cluster), and $\beta$ is the power law index.  The electron
temperature $T_e$ is taken to be a constant.  

As discussed in Section \ref{section:intro}, the X-ray surface
brightness from thermal bremsstrahlung radiation is given by
\begin{equation}
b_X(E)=\frac{1}{4\pi(1+z)^3}\int n_e^2(r)\Lambda(E,T_e)\,dl
\label{eq:XSB}
\end{equation}
\citep[e.g.,][]{Birkinshaw99}.  Under the assumption of isothermality
in the cluster, $\Lambda(E,T_e)$ is a constant.  Assuming spherical
symmetry and substituting Eq. \ref{eq:density}, this integral becomes
\begin{equation}
b_X \propto n_{e0}^2 \theta_0 D_A \left(1+\frac{\theta^2}{\theta_0^2}\right)^{-3\beta + 1/2}.
\label{eq:XSB2}
\end{equation}
Because surface brightness is independent of distance, and $D_A
\propto h^{-1}$, Eq.\ref{eq:XSB2} indicates that our measurement of
$n_{e0} \propto h^{1/2}$ (since $b_X$ and $\theta_0$ are constant).

The SZE signal is given by
\begin{equation}
\Delta I_{\rm SZE} \propto  T_e \int n_e \, dl.
\label{eq:dioveri_int}
\end{equation}
Again substituting Eq. \ref{eq:density}, this integral becomes
\begin{equation}
\Delta I_{\rm SZE} \propto T_e n_{e0} \theta_0 D_A
\left(1+\frac{\theta^2}{\theta_0^2}\right)^{-\frac{3}{2}\beta +
\frac{1}{2}}.
\label{eq:dioveri_int2}
\end{equation}
Given the above dependences on $h$ of $n_{e0}$ and $D_A$, one can see
that the SZE intensity, $\Delta I_{\rm SZE} \propto$ \hminhalf.
Therefore, if the density profile and electron temperature can be
obtained from X-ray observations and the SZE decrement can be
measured, one can determine the Hubble constant.  $\Delta I_{\rm SZE}$
is our main ``observable,'' and \hminhalf is the quantity we obtain
from each individual cluster measurement.  For the most part, the main
sources of error are symmetric (and approximately Gaussian) in
\hminhalf, but not in $h$.  Since $h$ has a non-linear relationship to
our observable, we cannot average individual values of $h$ for the
sample and obtain meaningful errors.  Instead, we average together
\hminhalf from the individual clusters to get a sample value of
\hminhalfnospace, which we then convert to $h$ for the final
measurement.

\subsection{Determining cluster parameters from X-ray data}
We combine density profile results from \rosat \citep{MM2000} with
temperature measurements from \asca \citep{MFSV98,White2000} and
\bepposax \citep{DM2002}.  The \rosat PSPC has a spatial resolution of
30$''$ FWHM and a field of view of 1.5 degrees in diameter, which
makes it well-suited for observations of the low-redshift clusters in
our sample.  At a redshift of 0.05, the field of view corresponds to
3.7 $h^{-1}$ Mpc, which is significantly larger than the expected
virial radius for the clusters.  Due to its small energy range
(0.1--2.4 keV) and spectral resolution, it is not possible to obtain
sufficiently accurate temperatures from \rosat data.  \XMM and
\chandra are ideal for determining both accurate density and
temperature profiles, and we will combine those spectral imaging data
with our SZE observations in a future paper.  Here, we use published
data from \asca and \bepposax observatories, which both have energy
ranges from 1--10 keV, making them useful for determining temperatures
of the hot gas in galaxy clusters.

\subsubsection{Cluster Density Profiles}
Mason \& Myers (2000, hereafter MM2000) derived density profiles for 14
of the clusters in our sample using archival \rosat PSPC data.  The
parameters $\beta$ and $\theta_0$ can be derived by fitting a
$\beta$-model surface brightness profile to the X-ray observations:
\begin{equation}
I(\theta) = I_0 \left(1+ \frac{\theta^2}{\theta_0^2}\right)^{-3\beta+1/2}.
\label{eq:sbprof}
\end{equation}
The $\beta$-model density normalization, $n_{e0}$, can be calculated
from the total X-ray flux measured over the observed bandpass; see
MM2000 for details.  Table \ref{tbl:param} lists the best fit model
parameters from MM2000 which we use in our SZE analysis.  MM2000
present 2 different models for clusters which appear to have a cooling
flow.  In their primary models, the X-ray emission is fit with a
$\beta$-model component plus a gaussian for the cooling flow.  In
their alternate models, the central region of cluster emission is
excised to remove contamination from the cooling flow, which because
of its compact size, contributes negligibly to the SZE analysis.  We
present results using the MM2000 primary models here, but we note that
the final results change very little ($<0.3\sigma$) when the
alternate models are used.  Please see MM2000 for details of the X-ray
modelfitting.  Note that MM2000 assumed $q_0=1/2$, and in some cases,
they used slightly different redshifts and electron temperatures from
what we assume in this paper.  To account for these differences, we
have recalculated the $n_{e0}$ normalization values using the method
described in MM2000.  We used redshifts from the compilation of
\citet{SR99}.

\subsubsection{Cluster Temperatures}
Cluster temperatures $T_e$ can be determined through spectral
modeling, where high-resolution X-ray spectra are fitted with a
thermal emission model for a low-density plasma in collisional
ionization equilibrium (i.e., the $mekal$ or Raymond-Smith models in
XSPEC), which has been absorbed by Galactic hydrogen.  If an X-ray
detector has sufficient spatial resolution, such as \XMM or \chandra,
temperature profiles can be measured by binning the spectral data into
different regions across the detector.  Here, we rely on published
\asca and \bepposax results.  \asca has an energy dependent PSF which
makes it difficult to obtain accurate temperature profiles from the
data.  Single emission weighted temperatures over the entire cluster
derived from \asca should be reliable, and we use these as estimates
for ``isothermal'' temperatures in our SZE analysis.  \bepposax has
good spatial resolution (1$'$) and a better understood PSF, making it
a better candidate for determining temperature profiles.  \bepposax
results are available for 2 clusters whose results are reported here.
\citet{DM2002} report average temperatures for the clusters, excluding
cooling flow regions.  Where available, we combine these with the
\asca results, and in Section \ref{section:tprof} we use their mean
profile results to determine the magnitude of error we might expect
from our isothermal assumption.

Table \ref{tbl:temps} lists temperature results obtained from \asca
data by \citet{MFSV98} and \citet{White2000}.  The results are in fair
agreement, except for the cooling flow clusters, where assumptions
used in the energy-dependent PSF matter more.  Until we can
conclusively resolve these discrepancies using observations from
current missions, we adopt the average of the results from both these
papers.  Given that the results are based on the same observational
data, the errors are likely to be correlated.  As a conservative
estimate of the error in the mean of the temperatures, we use the
larger of the two sets of error bars, and we include an extra
component to the uncertainty to account for systematic offsets between
the two measurements.
Note that \citet{MFSV98} quote 90\% confidence errors, while
\citet{White2000} uses 68\% confidence intervals.  For consistency, we
convert \citet{MFSV98} errors to 68\% confidence.  Since we do not
have knowledge of the actual likelihood distribution, we assume a
Gaussian distribution, symmetrize the errors, and scale them by 1.65.
Two of the clusters here have \bepposax observations which have been
analyzed in detail \citep{DM2002}.  These results are independent of
the \asca temperatures, and Table \ref{tbl:temps} shows them to be in
excellent agreement.  Where available, we average the \bepposax
temperatures with the mean \asca temperatures.  The temperatures and
errors we assume are listed in Table \ref{tbl:temps}.

\subsection{Modeling the expected SZE profile}
\label{section:szeprofile}

Including relativistic effects, the thermal SZE for an isothermal
cluster can be represented by
\begin{equation}
\frac{\Delta I_{\rm SZE}}{I} = \tau \frac{x e^x}{e^x-1} \left\{
  \frac{k_bT_e}{m_ec^2} (F - 4) + \left(\frac{k_bT_e}{m_ec^2}\right)^2
  \left[-10 + \frac{47}{2}F - \frac{42}{5}F^2 + \frac{7}{10}F^3 +
  \frac{7}{5}G^2(-3+F)\right] \right\}
\label{eq:sz1}
\end{equation}
\citep{SS98,CL98}.  $I$ is the CMB intensity, $\Delta I_{\rm SZE}$ is
the change due to the SZE, $k_b$ is the Boltzmann constant, $m_e$ is
the electron mass, and $c$ is the speed of light; $\tau$ is the
optical thickness to Compton scattering given by $\tau = \int \sigma_T
n_e(l) dl$, and $\sigma_T$ is the Thomson scattering cross section;
$x$ represents the frequency of observation, scaled as $x =
\frac{h\nu}{k_bT_{CMB}}$, with $\nu$ being the observing frequency and
$T_{CMB} = 2.725 \pm 0.001$ K \citep{T_COBE}; $F=x \coth(x/2)$ and $G=
x/\sinh(x/2)$.  The first term in Eq. \ref{eq:sz1} ($\propto
\frac{k_bT_e}{m_ec^2}$) represents the original thermal SZE described
by \citet{sz2}.  \citet{Reph95} showed that the relativistic
velocities of electrons in the hot gas of galaxy clusters must be
taken into account when measuring the SZE.  The second term in
Eq. \ref{eq:sz1} ($\propto \left(\frac{k_bT_e}{m_ec^2}\right)^2$) is
the relativistic correction \citep{SS98, CL98}.  This analytical
expression for the correction has been shown to be in good agreement
with numerical results of \citet{Reph95}.  For cluster gas with
temperatures in the range of our sample ($T_e \sim 4 - 10$ keV), the
relativistic term amounts to $\sim$3\% downward correction in the
magnitude of the predicted SZE at our frequency of observation.

We can factor out one $\left(\frac{k_bT_e}{m_ec^2}\right)$ and rewrite
Eq. \ref{eq:sz1} as:
\begin{equation}
\frac{\Delta I_{\rm SZE}}{I} =  \frac{k_bT_e}{m_e c^2} \sigma_T
f(x,T_e) \int n_e \, dl
\label{eq:dioveri_int3}
\end{equation}
where $f(x,T_e)$ is $\frac{x e^x}{e^x-1}$ multiplied the expression in
brackets (divided by $\frac{k_bT_e}{m_ec^2}$) from Eq. \ref{eq:sz1}.
From Eq.\ref{eq:density} we obtain the expected SZE profile,
\begin{equation}
\frac{\Delta I_{\rm SZE}}{I}(\theta) = \frac{k_b T_e}{m_ec^2}\sigma_T f(x,T_e) n_{e0}
\sqrt{\pi}\frac{\Gamma(\frac{3\beta}{2}-\frac{1}{2})}{\Gamma(\frac{3\beta}{2})}
D_A \theta_0 \left(1+\frac{\theta^2}{\theta_0^2}\right)^{-\frac{3}{2}\beta+\frac{1}{2}}.
\label{eq:dioveri}
\end{equation}
As we discuss later, the CMB contamination is sufficiently large that
we cannot accurately determine the $\beta$-model parameters from the
CBI data.  The parameters are, however, very well constrained by the
\rosat imaging observations, and we hold the model parameters fixed.
For each CBI frequency channel, Eq. \ref{eq:dioveri} can then be
reduced to
\begin{equation}
\Delta I_{\rm SZE}(\theta) = I_0\left(1+\frac{\theta^2}{\theta_0^2}\right)^{-\frac{3}{2}\beta+\frac{1}{2}},
\label{eq:dioveri2}
\end{equation}
where $I_0 (\propto h^{-1/2})$ is a constant.

\begin{figure*}
\plotone{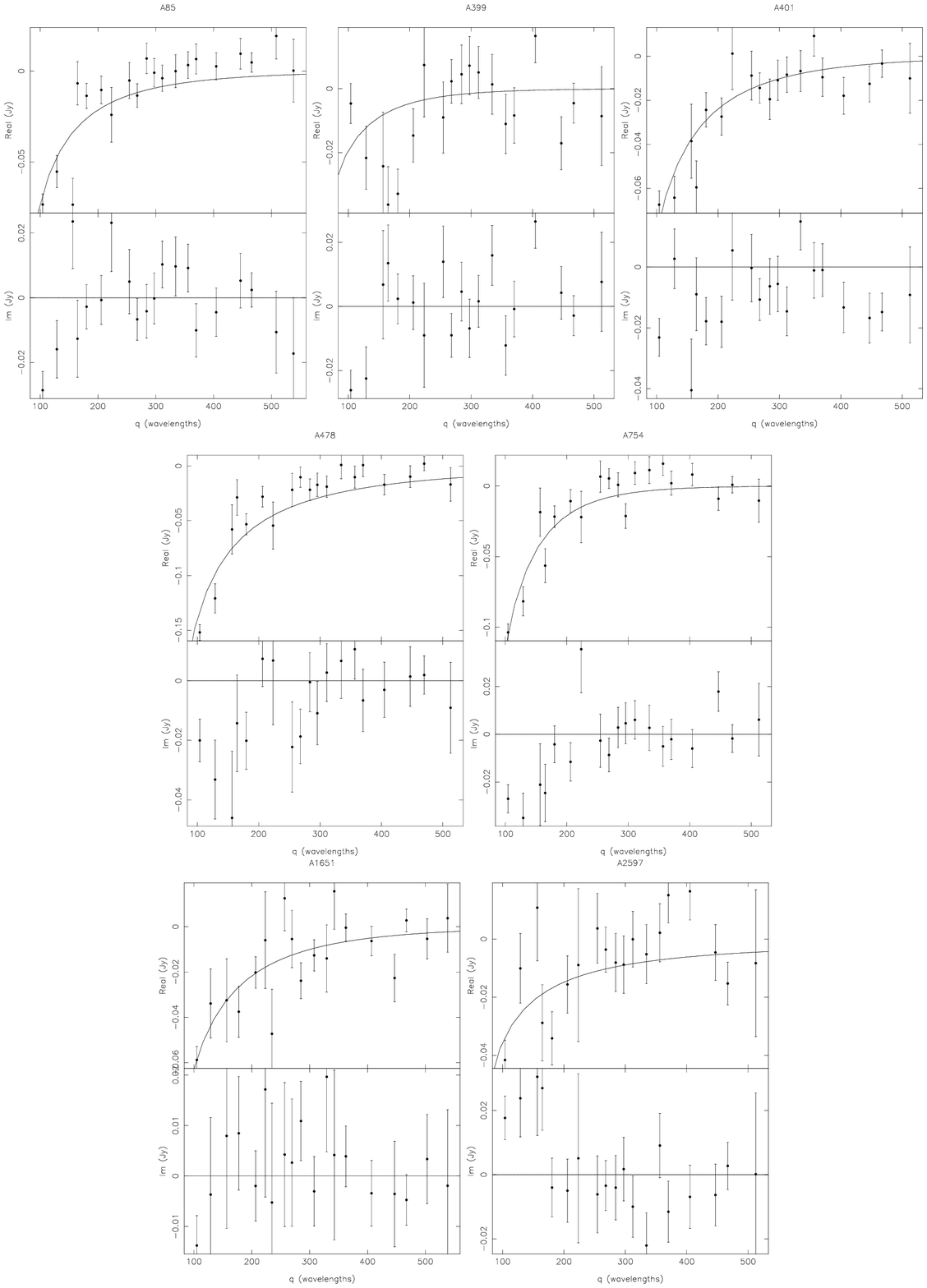}
\caption{Real and Imaginary visibilities showing radially averaged CBI
data with best fit model profiles. The radial length of the visibility
in wavelengths is given by $q= \sqrt{u^2+v^2}$.  At 30 GHz, 100
wavelengths = 1 m.}
\label{fig:visprof}
\end{figure*}

An interferometer measures the Fourier transform of this profile
multiplied by the primary beam of the telescope:
\begin{equation}
V(u,v) = I_0\int_{-\infty}^{\infty}\int_{-\infty}^{\infty}B(\theta)
\left(1+\frac{\theta^2}{\theta_0^2}\right)^{-\frac{3}{2}\beta+\frac{1}{2}}
e^{2\pi i (ux+vy)} \, dx \, dy,
\label{eq:vis}
\end{equation}
where $x$ and $y$ are positions on the sky ($\theta^2 = x^2 + y^2$),
and $u$ and $v$ are the visibility positions in units of wavelength.
Details of the CBI primary beam, $B$, are presented in
\citet{CBIPaper3}.  We fit the visibility model in Eq. \ref{eq:vis} to
the observed CBI data by minimizing $\chi^2$ with respect to $I_0$ to
obtain the best-fit SZE decrement and \hminhalfnospace.  The best-fit
visibility profiles are plotted with the radially averaged, point
source subtracted CBI data in Figure \ref{fig:visprof}.  Table
\ref{tbl:fit_results} lists results from the fits to the CBI
observations in mJy arcm$^{-2}$ and gives the $\chi^2$ values for the
fits.

In the context of interferometer observations, it is convenient to use
intensity units of Jy sr$^{-1}$, but more traditional single dish
observations quote SZE decrements in $\mu$K.  We use
\begin{equation}
\Delta I_{\rm SZE} = \frac{2\nu^2 k T_{CMB}}{c^2}\frac{x^2 e^x}{(e^x-1)^2} \frac{\Delta T}{T_{CMB}}
\end{equation}
to convert from intensity to $\mu$K.  
Table \ref{tbl:final_results} lists results from the fits to the CBI
observations in $\mu$K, and gives the $\chi^2$ values for the fits.
Another useful quantity is the Compton-$y$ parameter, defined as
\begin{equation}
y = \frac{k_bT_e}{m_ec^2}\tau,
\end{equation}
which is independent of the frequency of observation.  We list $y_0$,
the central Compton-$y$ value for each cluster in Table
\ref{tbl:final_results}.

\section{Error Analysis}
\label{section:errors}

The analysis method described above makes several idealizing
assumptions about galaxy clusters - that they are spherical, smooth,
and isothermal.  In this section we discuss possible implications of
deviations from these assumptions.  So far, we also have not
considered effects of contaminating factors such as observational
noise, CMB primary anisotropies, foreground point sources, non-thermal
radio emission from relics or haloes, and kinematic SZE signals from
peculiar velocities.  We address all these different sources of error
in this section.  Since the SZE model-fitting is performed in the
visibility domain (the Fourier transform of the image plane), the
error sources have relationships which are not analytical and whose
interpretations are not always intuitive.  Therefore, to characterize
their effects, we use Monte Carlo simulations, mimicking the real
observations and various error sources.

In the simulations, we attempted to reproduce as accurately as
possible all the components that enter a real CBI observation.  We
derived an SZE model ``image'' using the cluster gas parameters
obtained from the X-ray data as described in Section
\ref{section:szeprofile}.  We multiplied the image by the CBI primary
beam, and performed a Fourier transform to obtain our simulated model
visibility profile.  We used the observed CBI visibility data as a
template, maintaining identical $u-v$ coverage to the real observation
by replacing the observed visibility data with the simulated data and
randomizing the visibilities with the observed level of Gaussian
thermal noise.  We then analyzed each mock data set in the same way as
the actual observation, fitting for the ``observed'' SZE decrement.
We repeated this process $10^3$ times for each cluster, randomizing
the thermal noise and the error source whose impact we were attempting
to quantify.  This yielded a distribution of best-fit $I_0$'s, which
is equivalent to the distribution of \hminhalf for that error source,
which we use to obtain 68\% confidence intervals.
As we discuss below, several sources of error are not independent, and
must be considered together in the Monte Carlo simulations.  The
largest source of random error is the intrinsic CMB anisotropy.
It has a significant impact on almost all the other error sources, so
we include it in most of the other simulations.

\subsection{Intrinsic CMB Anisotropies}
\label{section:cmb}

The CBI has measured the CMB on arcminute scales, finding bandpower
levels of 2067$\pm$ 375 $\mu$K$^{2}$ at $\ell\sim600$ (1 m baseline),
and 1256$\pm$ 284 $\mu$K$^{2}$ at $\ell\sim1200$ (2 m baseline)
\citep{CBIPaper3}.  Figure \ref{fig:visprof} shows that the SZE
cluster signal is strongest on the 1 m and 2 m baselines, where the
CMB is a significant contaminant.  The SZE data is effectively
radially averaged in the visibility fitting, and the rms of the CMB
averaged in this way on the 1 m and 2 m baselines is 30 mJy and 7 mJy,
respectively.  We cannot remove the intrinsic CMB anisotropies from
our data without observations at other frequencies, so we need to
measure its impact on our results.  We generated $10^3$ randomized
realizations of the CMB primary anisotropies, using the algorithm
described in Appendix A.  Each of these ``sky'' realizations was then
added to the simulated clusters described above, and we fitted for the
value of \hminhalf which minimized $\chi^2$.  The input power spectrum
we used is the best fit model to the CBI power spectrum observations,
combined with Boomerang-98, DASI, Maxima, VSA, and {\em COBE} DMR
measurements \citep{CBIPaper5}:$\Omega_{\rm tot}=1.0$, $\Omega_{b}h^2
= 0.02$, $\Omega_{\rm cdm}h^2=0.14$, $\Omega_\Lambda= 0.5$,
$n_s=0.925$, $\tau_c=0$, ${\cal C}_{10} = 887$ $\mu$K$^2$.  We list in
Table \ref{tbl:errors} the 68\% confidence intervals in \hminhalf for
each cluster, given the expected levels of CMB contamination based on
the CBI's power spectrum measurements.  The average fractional error
in \hminhalf per cluster due to the CMB is 36\%, and this clearly
dominates all other sources of uncertainty.

\subsection{Density model errors}

Because the CMB contamination is so large, it is not meaningful to fit
for the shape of the cluster gas profile from the SZE data.  We
therefore assume that the profile we derived from the X-ray data is
correct, and hold the $\beta$-model parameters fixed.  Here, we
quantify errors due to possible deviations from this best fit X-ray
model.  To determine the error in the individual cluster density
profile parameters MM2000 also used Monte Carlo simulations.  For each
cluster, they smoothed the original composite 0.5-2.0 keV count-rate
image using a 30$''$ FWHM Gaussian.  A set of $10^3$ simulated
observations were then created by multiplying the smoothed image by
the exposure maps and adding random Poisson noise.  For each simulated
observation, they applied the same analysis procedure that was used to
determine the cluster parameters from the original data set.  We use
their resulting distribution of $\beta$-model parameters, $\beta$,
$\theta_0$, and $n_{e0}$ to determine the expected error in \hminhalf
due to possible ambiguities in the density profile modeling for each
cluster.  We generated $10^3$ simulated CBI data sets using the
different $\beta$-model parameter trios from the simulated X-ray
observations to generate slightly different SZE profiles.  We then fitted
for the SZE decrement $\Delta I$ using the best fit X-ray parameters
from the original \rosat image.  The resulting distribution in
$h^{-1/2}$ provides expected errors due to possible inaccuracies in
the density profile modeling.  Because the X-ray emission and SZE
have different dependences on the model parameters, there is a slight
bias in the SZE distribution relative to the X-ray distributions.  We
discuss this bias in Appendix \ref{section:bias} and list the results
in Table \ref{tbl:errors}.  The bias corrections are mostly
negligible ($<$ 1\%), but A754, a highly disturbed cluster with a larger degree
of model parameter uncertainty, requires a correction of 3.6\%.

\subsection{Temperature Profiles}
\label{section:tprof}
Our analysis also assumes that cluster gas is isothermal.  If this
assumption is correct, determining errors from inaccuracies in the
value of $T_e$ is straightforward; \hminhalf is simply proportional to
$T_e$.  Whether the gas is in fact isothermal has been the subject of
ongoing debate.  In their analysis of the same \asca data,
\citet{MFSV98} find temperature profiles which decline with radius,
while \citet{White2000} finds isothermal profiles.  \citet{DM2002}
also find declining profiles from their analysis of \bepposax data,
but they find that the profiles have a slightly different slope and
break radius from the \citet{MFSV98} profiles.  \XMM observations
indicate that individual clusters may vary; A1795 has a temperature
profile consistent with isothermal out to 0.4 $r_{\rm vir}$ while Coma
shows a declining temperature profile \citep{xmm_a1795,xmm_coma}.
Departures from isothermality can produce large errors in the
derivation of $H_0$ from the X-ray/SZE method if an isothermal model
is assumed, but the magnitude of the error depends on many factors.

To estimate the possible effect of an inaccurate temperature profile,
we study the case of gas modelled as a hybrid isothermal-polytropic
temperature profile, where the temperature is uniform out to a radius
$r_{\rm iso}$, and declines outside this radius.  This model was
introduced by \citet{hybrid_tprof} and is similar to the profile found
by \citet{NVW95} in N-body simulations.  It is represented by
\begin{equation}
T(r) = \left\{ \begin{array}{ll} 
		T_0 & \mbox{if }r\leq r_{\rm iso} \\
                T_0\left(\frac{n(r)}{n(r_{\rm iso})}\right)^{\gamma-1} & \mbox{if } r>r_{\rm iso}
	     \end{array} \right. .
\end{equation}
Theoretical calculations disagree on where the transition radius
$r_{\rm iso}$ should occur, although it is generally taken to be of
order a virial radius, $r_{\rm vir}$, which we approximate as
$r_{200}$, defined in the manner of \citet{EMN96} as the radius which
encloses a mean density 200 times the cosmological critical density.

In the limit where $r_{\rm iso}=0$, the temperature profile is simply
a polytropic model.  We expect $1<\gamma<5/3$, where $\gamma=1$ is the
isothermal limit, and $\gamma=5/3$ is the adiabatic limit.  The
expected central SZE decrement depends fairly strongly on $\gamma$ and
$r_{\rm iso}$, but the effect of these parameters on the derivation of
$H_0$ using interferometric SZE data is not completely
straightforward.  A hybrid temperature profile will cause two changes
relative to an isothermal model.  The overall decrement will be
smaller, and the cluster will appear more compact.  The interferometer
measures visibilities which are the Fourier transform of the image, so
a steep image profile will have a shallower visibility profile and
vice versa.  Because the visibility profile shallows as $r_{\rm iso}$
is decreased, the hybrid profiles cross the isothermal profile at
different points.  Therefore, it is difficult to know whether we will
overpredict or underpredict $H_0$ for different clusters without an
accurate temperature profile.

We demonstrate this with a simulation which is summarized in Table
\ref{tbl:a478tprof}.  We generated false SZE data sets using hybrid
models with different pairs of $\gamma$ and $r_{\rm iso}$ where the
input models all assumed $h=1$.  For each model, we rescaled $T_0$
such that the emission weighted temperature for that model agreed with
the observed value.  We then fitted an isothermal profile to the
hybrid model data to determine the error in deriving $H_0$ due to the
temperature profile.  Table \ref{tbl:a478tprof} lists our results for
A478.  We see that for steeply declining profiles, our error in $h$
will be very large, and can be incorrect by a factor of 2 in the most
extreme case.  Some cases supported by observational data include
$r_{\rm iso}=0$, $\gamma=1.2$ \citep{MFSV98} and $r_{\rm iso}=0.2$,
$\gamma=1.5$ or $\gamma=1.2$ for non-cooling flow and cooling flow
clusters respectively \citep{DM2002}, where $r_{\rm iso}$ is in units
of $r_{200}$.  If the \citet{MFSV98} profile is correct, then we
overestimate $h$ from A478 by 22\%; if the \citet{DM2002} profile is
correct, then our value of $h$ is largely unaffected by the
temperature profile.  The errors can be very different for different
clusters, and in some cases go in the opposite direction, where $h$ is
underestimated for steeply declining temperature profiles.  We
summarize in Table \ref{tbl:tprofsum} the levels of error expected if
the mean \asca and \bepposax profiles apply to each of our clusters.
For the \citet{DM2002} profiles, we have boldfaced the relevant
column, based on whether a particular cluster is believed to have a
cooling flow or not.  Depending on which mean profile is assumed, for
the sample of 7 clusters presented here, our value of $h$ may be
essentially correct, with an overestimate of only 1\% based on the
\citet{MFSV98} profile, or it could be underestimated by 14\% assuming
the \citet{DM2002} mean profiles.  This demonstrates that an accurate
knowledge of the temperature profile is important for eliminating a
bias in $H_0$ from non-isothermal cluster temperatures.  To determine
$r_{\rm iso}$ and $\gamma$, the profiles need to be probed to a large
radius, typically tens of arcminutes for our clusters, which means
that \chandra observations alone are usually not adequate for these
purposes.

\subsection{Errors from Foreground Point Sources}
\label{section:ptsrc}

Foreground point sources are the largest source of contamination in
CMB experiments at 30 GHz.  We hope to limit the point source error in
our $H_0$ determination to $<2\%$ for the sample of 15 clusters, or
$<8\%$ per cluster.  Our strategy for removing point source
contamination involves a combination of fitting for the fluxes of
sources at known positions simultaneously with the cluster model, and
independently measuring some source fluxes with the OVRO 40-m
telescope and the VLA.  The CBI short ($\leq$2 m) baselines are most
sensitive to the signal from the extended cluster and CMB primary
anisotropies.  The strength of the cluster and CMB both decline
significantly on longer baselines, making those baselines suitable for
determining point source fluxes.  However, this source fitting is not
reliable for sources near the cluster center, and we use independent
observations to accurately determine their fluxes.  The advantage of
fitting the source fluxes using the CBI data is that the point source
and cluster observations are simultaneous, so source variability is
not an issue, although most point sources in clusters are steep
spectrum and non-varying \citep[e.g.][]{slingo74,cooray_pt}.   The
disadvantage is that for sources close to the cluster center (also the
pointing center of the observation), the point source appears as an
overall offset on all baselines in the visibility domain where we
perform the fitting.  The overall offset from the point source is
difficult to distinguish from the cluster signal, especially in the
less resolved, compact clusters such as A478 and A401, resulting in
large errors in the $H_0$ analysis.  We find that sources close to the
cluster center within about 10$'$ need to be observed independently
with very high accuracy (about 1 mJy rms at 31 GHz), and sources
outside this radius can be safely fitted using the CBI long baselines.
We fit sources outside the 10$'$ radius that have fluxes detectable at
the 2.5-$\sigma$ limit, and we account for the contribution of the
remaining (unfitted and unsubtracted) sources statistically using
Monte Carlo simulations which we describe below.  As a basis for our
study, we use the NVSS catalog, which is complete to 2.5 mJy at 1.4
GHz.  We assume that all relevant sources at 30 GHz are in the NVSS
catalog and that we do not miss any sources with inverted spectral
indices.  This assumption is supported by our OVRO study and VLA
X-band survey of one of the blank CMB fields, as well as a study of
point sources in the CBI Deep fields \citep{CBIPaper2}.  Although
\citet{taylor_pt} do find from their 15 GHz survey a large chance of
missing inverted spectrum sources from extrapolations from low
frequency, we estimate based on their source counts that the
probability of such a source occuring in the crucial central 10$'$ of
our cluster centers is low.  Sources outside this radius do not
contribute a significant error, as we explain below.

To test our source subtraction method and quantify errors, we
generated simulated cluster observations using the X-ray derived
density models described above, including realistic thermal noise and
CMB primary anisotropies.  We added all NVSS sources (down to 2.5 mJy
at 1.4 GHz) to each simulated cluster realization at the listed NVSS
positions, but we varied the fluxes for each iteration by randomly
selecting spectral indices from the distribution observed by
\citet{CBIPaper2} to determine the source fluxes at each of the CBI
channels from 26 to 36 GHz.  We then defined specific criteria to
determine how the different sources would be treated in the analysis.

If a source was within 10$'$ of the LMT pointing centers, we
``subtracted'' the source, assuming its flux is known to a certain rms
from an independent telescope.  In the simulations we added at these
source positions, random Gaussian noise with an rms equivalent to the
levels observed with the OVRO 40-m telescope.  The sensitivity
achieved with the 40-m varied from 0.4 to 2.0 mJy rms for individual
sources.  
The simulations show that these central point sources
observed with the 40-m contribute $\sim$ 10\% error per cluster.

If a source was outside the 10$'$ radius and could be detected at the
2.5-$\sigma$ level on baselines longer than 2.5 m, we fitted for the
flux of the source simultaneously with the cluster.  Before selecting
the 2.5-$\sigma$ sources in the simulations, we added random
fluctuations to the source fluxes to simulate possibly missing some
sources due to noise.  All other sources were ignored (i.e., not
subtracted or fit for in any way).  In the fitting, we fixed the
source positions to the NVSS coordinates, which is reasonable since
the rms error in the NVSS positions ranges from $<1-7''$, much smaller
than the CBI synthesized beam of a few arcminutes.
We found that it was extremely inefficient to fit for the spectral
indices of large groups of point sources over the CBI 10 GHz
bandwidth, so we fixed them at the weighted mean value of the
distribution found by \citet{CBIPaper2}, $\alpha=-0.55$.  To determine
whether this assumption affects our results, we also tried fixing the
spectral index to $\alpha=0$.  We found that the $H_0$ results change
by $<1$\% in all cases.
In the simulations the small number of sources whose fluxes are
determined from the CBI data itself (typically 10-15 per cluster)
contribute a negligible amount of error to the \hminhalf fits.

All sources that were outside the 10$'$ radius and were not at least
2.5-$\sigma$ were ignored in the SZE fitting.  We compare these
simulations with those described in Section \ref{section:cmb}, where
only observational noise and CMB anisotropies are added.  The $10^3$
Monte Carlo iterations show that the unsubtracted sources contribute a
small but consistent bias, tending to make the mean \hminhalf for a
cluster higher or lower, depending on the configuration of residual
sources present in the LMT fields.  The bias from the unsubtracted
sources is an additive factor, and the values for the individual
clusters are listed in Table \ref{tbl:errors}.  All are $<$ 3\%.

The spectral index distribution determined by \citet{CBIPaper2} was derived
for observations of cluster-free CMB fields.  The point source
populations in galaxy clusters can be considerably different, and have
not been well studied at 30 GHz.  \citet{cooray_pt} find a distribution of
$-0.77\pm 0.48$, which is somewhat different from the \citet{CBIPaper2}
distribution.  We retest our method using the \citet{cooray_pt}
distribution instead of the \citet{CBIPaper2} distribution, and find the
results to be almost unchanged.  The sample value of \hminhalf changes
by $<$0.3\%, and the magnitude of the sample error changes by $<$1\%.

\subsection{Other error sources}

\subsubsection{Asphericity}
Because of the CMB contamination, we cannot meaningfully study the
cluster shapes as seen in the SZE from the CBI data.  Any errors in
\hminhalf from slight pointing inaccuracies on the level seen in the
CBI, offsets in $x,y$ from the assumed cluster center up to a few
arcminutes, or ellipticity in the plane of the sky are all dwarfed by
the CMB, so we ignore them here.  The 2-D shapes of clusters seen in
X-ray emission provide a good indicator of the level of expected
asphericity.  \citet{Cooray2000} has analyzed the 2-D distribution of
X-ray cluster shapes observed by \citet{einstein_morph2}, showing that
for a sample of 25 clusters randomly drawn from an intrinsically
prolate distribution, the error in $H_0$ for the sample is less than
3\%.  Therefore, for our complete sample, we do not expect a large
systematic error due to cluster asphericity.  We estimate the
uncertainty in $H_0$ for each cluster by taking 3\% $\times \sqrt{25}$
= 15\%, so the uncertainty for each cluster in \hminhalf $\sim 7.5$\%.
For our primary sample of 15 clusters, the error in $H_0$ due to
asphericity should be $<$4\%.

\subsubsection{Clumpy gas distribution}

In the fitting, we assume that the density distribution is smooth,
directly applying the density profile model derived from the X-ray
observations to the SZE models.  Because the CMB is such a large
contaminant, and we cannot meaningfully obtain shape parameters from
the SZE data, potential systematic errors due to clumpy gas cannot be
addressed with our data and can probably only be understood in detail
through hydrodynamical simulations.  However, Eq. \ref{eq:XSB} and
Eq. \ref{eq:dioveri_int2} show that $h \propto \langle n_e^2 \rangle/
\langle n_e \rangle ^2$, a quantity which is always greater than
unity.  Therefore, any clumpiness in the gas distribution will cause
one to overestimate $h$ by this factor, although as we saw from our
study of different temperature profiles that the analysis for
interferometer data could be more complicated than this.

\subsubsection{Peculiar Velocities}

The expression given for the thermal SZE in Eq. \ref{eq:sz1} assumes that
the cluster is not moving with respect to the Hubble flow.  In
reality, all clusters have some peculiar velocity $V_{\rm pec}$, taken to be at
an angle $\theta$ relative to the vector drawn from
the cluster to the observer.  This produces a kinematic SZE
given by
\begin{equation}
\frac{\Delta I_{\rm kin}}{I} = \tau \frac{x e^x}{e^x-1} \left \{
  \frac{V_{\rm pec}}{c}\mu + \left(\frac{V_{\rm pec}}{c}\right)^2\left(-1-\mu^2 +
  \frac{3+11\mu^2}{20}F\right) +
  \frac{V_{\rm pec}}{c}\frac{k_bT_e}{m_ec^2}\mu\left[10 - \frac{47}{5}F +
  \frac{7}{10}(2F^2 + G^2)\right]\right\}
\label{eq:kin_sz}
\end{equation}
\citep{SS98}, where all quantities are as defined in Section
\ref{section:szeprofile}, and $\mu=\cos\theta$.  The first term is the
kinematic SZE, which can be positive or negative, depending on whether
the cluster is moving away from us or towards us.  The second term is
a relativistic correction to the kinematic SZE, and the third part of
the expression is a cross-term between the thermal and kinematic
effects.  The last term is the dominant correction to kinematic SZE
measurements in single clusters.  For our purposes, we can assume that
the clusters have random peculiar velocities along our line of sight,
and the terms in $V_{\rm pec}/c$ will average out over the sample, although we
will consider their contribution to our error budget for each
individual cluster below.  The $(V_{\rm pec}/c)^2$ term will not average out for
the sample, but its magnitude is very small, even for large peculiar
velocities.

\citet{pec_v} measured the peculiar velocities of 24 galaxy clusters
with radial velocities between 1000 and 9200 km s$^{-1}$ ($z<0.03$).
None of the peculiar velocities in the CMB reference frame exceed 600
km s$^{-1}$, and their distribution has a line of sight dispersion of
300 km s$^{-1}$.  The mean magnitude of their observed radial peculiar
velocities is $\sim$200 km s$^{-1}$.  Slightly larger peculiar
velocities are found in numerical simulations created by the Virgo
Consortium \citep[e.g.,][]{virgo_sim}.  \citet{colberg} find from the
simulations that for haloes with masses comparable to the clusters in
our sample ($M>3.5 \times 10^{14} h^{-1}$\msun), the peculiar
velocities range from about 100$-$1000 km s$^{-1}$ for a $\Lambda$CDM
cosmology, with 11 out of 69 clusters having $V_{\rm pec}>600$ km
s$^{-1}$.  The average peculiar velocity for the $\Lambda$CDM model is
about 400 km s$^{-1}$.  We calculate the error in \hminhalf due to the
kinematic SZE for each cluster assuming $V_{\rm pec} =$400 km s$^{-1}$
and list the values in Table \ref{tbl:errors}.  The uncertainties due
to peculiar velocities range from 4\%--8\%.  For the weakest cluster
in our sample, A2597, a large peculiar velocity of 1000 km s$^{-1}$
produces an error in \hminhalf of $\sim$0.04\% from the $(V_{\rm
pec}/c)^2$ term.  Thus, there is no systematic error in the sample
from ignoring the kinematic SZE relativistic correction.  Errors due
to the kinematic/thermal SZE cross term are negligible ($<0.1$\%) for
all cases.

\subsubsection{Comptonization due to a nonthermal populations of electrons}
\label{section:nt}
Clusters which have recently merged are usually associated with radio
relics or haloes.  This non-thermal emission can affect our results in
two possible ways.  If the halo is extremely bright, it can cause
contaminating foreground emission at 31 GHz.  As noted in the sections
on the individual clusters, two clusters discussed in this paper, A401
and A754, have haloes at lower frequencies, but due to their steep
spectral indices, we do not expect them to contribute significant flux
at 31 GHz.  The second possible effect is a contribution to the SZE
from the non-thermal population of electrons present in the cluster.
Quantifying this effect is very difficult, even with detailed radio,
EUV, and X-ray data because the electron population models are not
well-constrained by the observational data.  \citet{nt_sz1} have
studied four different electron population models that can reproduce
observed data from the Coma cluster and A2199.  Three of the four
models produce a negligible contribution to the total SZE from the
non-thermal population of electrons ($<1$\% in most cases).  The
fourth model non-thermal electron population contributes 6.8\% and
34.5\% of the total SZE flux in the two clusters, but that model is
deemed to be unviable by the authors because the ratio of the total
energy in the non-thermal electrons to that in the thermal electrons
is too high to be realistic.  One of the three viable models produces
a contribution of 3\% to the total SZE in A2199.
\citet{nt_sz2} also calculate the contribution to the SZE from
non-thermal electrons, using fewer approximations than \citet{nt_sz1}.
Their results are similar in that the magnitude of the non-thermal SZE
is highly dependent on the model used to represent the electron
population.
Here, we simply note that Comptonization by a non-thermal electron
population is a possible source of additional error in our result,
although current plausible models indicate that the magnitude of the
error may be negligible.

\section{Combined Results}
\label{section:results}

To determine the total error for each individual cluster, we can
combine independent sources of error by adding them in quadrature if
they are Gaussian, or convolving the different likelihood
distributions if they are not.  However, since all the Monte Carlo
simulations depend so heavily on the CMB noise, it is difficult to
separate the independent components.  For example, the CMB has a large
effect on how well the point sources can be fitted, as do the X-ray
models.  In the point source fitting, we assume we completely
understand the shape of the SZE profile based on the X-ray data.  If
the profile is slightly wrong, this can affect the point source
subtraction.  We therefore perform a final simulation set where all
the major sources of error - CMB, point sources, and X-ray models -
are varied simultaneously.  Here we only consider the isothermal
$\beta$-models described in MM2000.  
We list in Table \ref{tbl:errors} the 68\% confidence intervals in
\hminhalf for each cluster, given the various sources of error.
Note that these errors are absolute, not fractional.  The reason is
that the CMB dominates the error, and it effectively adds or subtracts
flux to the cluster and is not a scaling factor, although the
fractional expected error in $h^{-1/2}$ is a function of the cluster
signal and shape, where brighter, more compact clusters experience
less contamination.

The final \hminhalf results, corrected for the X-ray model and
unsubtracted point source biases, are presented in Table
\ref{tbl:final_results} with their 68\% random uncertainties from CMB
anisotropies, thermal noise, calibration errors, point source
subtraction, asphericity, temperature determination (within context of
an isothermal model), and peculiar velocities.  The final statistical
uncertainties were calculated by adding in quadrature the values from
the last three columns of Table \ref{tbl:errors}, the 7.5\% uncertainty
per cluster from asphericity, and the 2\% uncertainty per cluster
from the radio calibration.  Central values of $\Delta T_0$ and
$y_0$ which have also been corrected for the X-ray model and
unsubtracted point source biases are listed as well.

As we have emphasized, a major potential bias in determining $H_0$
from combined SZE/X-ray observations is asphericity in the clusters.
Weighting by the errors could possibly bias the sample average if the
magnitudes of the errors correlate with properties of the cluster that
relate to the asphericity bias.  For example, an elongated cluster
oriented along the line of sight would appear more compact than the
same cluster oriented perpendicular to the line of sight.  Thus, if
the magnitudes of the errors in our determination of $H_0$ correlate
with the sizes on the sky of the clusters, a sample result which has
been weighted by the errors could potentially be biased.  We determine
the apparent sizes of each cluster by calculating the FWHM of the
$\beta$-model.
The most compact cluster, A2597, is also one of the least luminous
objects in the sample.  Since its signal is relatively weak, the error
due to CMB contamination is significant.  If we exclude this cluster,
there is very little correlation between the errors and cluster size,
with a correlation coefficient $r=-0.16$; with this one anomalous
object, $r=-0.58$.

First, to avoid any possible bias, we simply take a straightforward
average of the $h^{-1/2}$ results and errors.  The unweighted average
is $H_0 = 67^{+30}_{-18}\mbox{}^{+15}_{-6}$ \hunit for this sample of
7 clusters, where the first set of uncertainties represents the random
error at 68\% confidence, and the second set represents systematic
errors corresponding to calibration uncertainties and possible bias
due to a nonisothermal profile, for which we use the average sample
biases ($^{+1\%}_{-8\%}$ in \hminhalf) from Table \ref{tbl:tprofsum}.
Since we expect the 5\% absolute calibration uncertainty to be
independent of the nonisothermality bias, we added the two
uncertainties in quadrature.  As discussed in Section
\ref{section:data}, A754 is a merging, disturbed cluster.  If we
exclude it from the final result, we obtain $H_0 =
65^{+34}_{-19}\mbox{}^{+14}_{-6}$ \hunit from the remaining 6
clusters.  Given that the correlation between error and cluster size
is not large, we also present a weighted sample average, with the
caveat that there is a possibility of the result being biased.  The
sample average weighted by the errors gives $H_0 =
75^{+23}_{-16}\mbox{}^{+16}_{-7}$ \hunit.  

We use the sample average value to compare the relative magnitudes of
the sources of statistical error discussed in Section
\ref{section:errors}.  In Table \ref{tbl:errors}, random errors from
the CMB and unsubtracted or incorrectly subtracted point sources are
given as absolute errors.  This is because the CMB and point sources
have a given strength and will cause the same magnitude of error
whether the cluster is weak or strong.  Random errors from
asphericity, electron temperature measurements, peculiar velocity, and
radio calibration are all fractional errors.  In Table
\ref{tbl:errsum}, we summarize all the sources of statistical
uncertainty, using our sample value of \hminhalf$=1.22$ to convert
between absolute and fractional uncertainties.  We give the average
error expected per cluster, as well as the sample average.

There is a large scatter in the individual \hminhalf results, but the
scatter is entirely consistent with the uncertainties.  The mean and
standard deviation in \hminhalf=1.22$\pm$0.52.  For the 7 clusters,
the error in the mean is 0.20, which is equal to the sample
uncertainty derived from the individual cluster errors of 0.20 from
the unweighted average.  The reduced $\chi^2$ for the sample mean is
1.47 with 6 degrees of freedom, with a probability of 21\% of
exceeding this value by chance.

The uncertainties in the $H_0$ results presented here are dominated by
confusion due to the CMB primary anisotropies.  In this analysis, when
fitting the SZE models to the visibility data, we weight only by the
thermal noise.  An obvious improvement would be to take advantage of
the fact that we know the CMB's angular power spectrum
\citep{CBIPaper2,CBIPaper3}, and weight the visibility data by the
level of power in the CMB on a particular angular scale when
performing the modelfitting.  However, the errors due to the CMB are
highly correlated for visibilities which are close to each other, and
this correlation must be removed by diagonalizing the CMB covariance
matrix.  Details of this method will be described in a future paper
\citetext{Udomprasert \& Sievers, in preparation}.  We note here that
by using this improved weighting method, the errors in our result
above would be reduced by $\sim 30$\% for this sample of 7 clusters.

\section{Comparison with Past SZE Observations}

\citet{MasonSZ} (hereafter MMR) and \citet{MyersSZ} observed four of
the clusters presented in this paper, A399, A401, A478, and A1651,
with the OVRO 5-m telescope.  We compare the CBI results with the OVRO
5-m observations.  MMR reanalyzed A478 observations taken by
\citet{MyersSZ}, and we use the MMR results here.  There are a few
differences between the CBI and OVRO 5-m observations which must be
taken into account.  First, for all 4 clusters, different Lead and
Trail fields were observed by the 2 groups.  These differing fields
contribute significant errors to the results.  Also, slightly
different redshifts, electron temperatures, and cosmologies were
assumed in the 2 analyses.  If we take these into account and fit
models to the CBI data using all the same parameters assumed by MMR,
the results we would obtain are presented in Table \ref{tbl:cf_mmr}.
Errors from the CMB in the Main fields will be correlated for the 2
observations, since the same patch of CMB is being observed.  However,
the CMB contribution should not be identical because the
interferometer and single dish measurements are sensitive to different
modes of the CMB.  Calculating the correlated error in the Main field
is complicated, so instead we performed the following estimate.  We
compared our results to those of MMR assuming two different
uncertainties.  In our first comparison, we included the entire 68\%
confidence errors as quoted in MMR, which included errors due to
contributions from the Lead, Main, and Trail fields, whereas for the CBI
measurements, we removed the contribution to the uncertainty from
the CMB in the Main field, but included uncertainties from CMB in the
Lead and Trail field, as well as thermal noise from the Main field.
In the second comparison, we 
removed the contribution
to the uncertainty from the Main field CMB in the MMR result as well.
Table \ref{tbl:cf_mmr} 
shows the results we obtain from these
comparisons.  We calculated $\chi^2$ to determine the probability due to
chance of our results differing by the observed amount.
\begin{equation}
\chi^2 = \sum \frac{(h^{-1/2}_{CBI} -  h^{-1/2}_{MMR})^2}{\sigma_{CBI,LT}^2 + \sigma_{MMR,L(M)T}^2}
\end{equation}
For the 4 clusters, we obtained $\chi^2/\nu$=1.53, for 4 degrees of
freedom, with an associated probability of 19\% when the Main field
CMB uncertainty is included once; $\chi^2/\nu$=2.43 when the Main
field CMB was ignored completely, with a probability of 5\%.  We expect
the actual value to be something between these, showing that the CBI
and OVRO 5-m results are in reasonable agreement.

\section{Discussion and Conclusions}

From the CBI's SZE observations of 7 low redshift clusters, we have
obtained a measurement of $H_0 = 67^{+30}_{-18}\mbox{}^{+15}_{-6}$
\hunit from an unweighted sample average.
We have quantified many sources of error, the largest being
contamination due to CMB primary anisotropies.  Observations of 12
more clusters have been taken, and their analysis will be published in
future papers.  In addition to the four clusters which we have studied
in common, MMR also determined $H_0$ from three additional clusters,
Coma, A2142, and A2256, which fall under our sample selection criteria
but are too far north to be observed with the CBI.  If we include
those three clusters in our sample average, we obtain a result of $H_0
= 68^{+21}_{-14}$ \hunit for an unweighted sample average, where the
quoted errors are random uncertainties at 68\% confidence.
The value of $H_0$ we obtain from the low redshift clusters is
entirely consistent with the value obtained by the {\em Hubble Space
Telescope} Key project of $H_0 = 72\pm8$ \hunit \citep{HSTKey} and the
{\em Wilkinson Microwave Anisotropy Probe} ({\em WMAP}) of $H_0 = 72\pm5$
\hunit \citep{WMAP}.  Our result is also consistent with that obtained
from the SZE at higher redshift by \citet{Reese2002} of $H_0 =
60^{+4}_{-4}\mbox{}^{+13}_{-18}$ \hunit, although our sample value is
somewhat higher.

In our current analysis, we perform straightforward fits to the simulated and
observed data, taking into account only the thermal noise as the
weighting factors in the fitting.  We expect to obtain significantly
improved results by a more refined treatment of the effects of intrinsic
anisotropy (see Section \ref{section:results}).
In future work, we will also attempt to address the errors due to
incorrect modeling of the cluster gas distribution.
\XMM and \chandra will provide definitive measurements of
temperature profiles for the clusters in our sample, and
hydrodynamical simulations will allow us to quantify errors from
clumpy, aspherical gas distributions.  
By making these improvements to our results, $H_0$ measurements from
the SZE will provide a powerful check on other methods, such as the
cosmic distance ladder of the HST $H_0$ Key Project.

\acknowledgments

We gratefully acknowledge Hans B{\" o}hringer and the REFLEX team for
sharing with us their cluster data in advance of publication.  We
thank Jonathan Sievers and Monique Arnaud for useful discussions, and
the anonymous referee for helpful suggestions that improved the
organization of the paper.  Steve Myers' early work in this field set
the stage for this project, and we acknowledge his assistance
throughout.  We are most grateful to Steve Padin, Martin Shepherd and
John Cartwright for their many invaluable contributions to the CBI.
We acknowledge the invaluable efforts of Russ Keeney, Steve Miller,
Angel Otarola, Walter Schaal, and John Yamasaki at various stages of
the project.  We gratefully acknowledge the generous support of Maxine
and Ronald Linde, Cecil and Sally Drinkward, Barbara and Stanley Rawn,
Jr., and Fred Kavli, and the strong support of the provost and
president of the California Institute of Technology, the PMA division
Chairman, the director of the Owens Valley Radio Observatory, and our
colleagues in the PMA Division.  This work was supported by the
National Science Foundation under grants AST 94-13935, AST 98-02989,
and AST 00-98734.  PSU acknowledges support from an NSF Graduate
Student Fellowship.

\appendix

\section{Simulations of CMB}
\label{section:cbisky}

\def\dir{\hat{\bf n}}
\def\EXP#1{\langle #1 \rangle}
\def\VAR#1{{\rm Var}( #1)}
\def\xvec{{\bf x}}
\def\uvec{{\bf u}}
\def\rvec{{\bf r}}
\def\sinc{\mathop{\rm sinc}\nolimits}
\def\half{{\textstyle{1\over2}}}

We generate simulated images of the CMB from an input power spectrum
using the method described below:

\begin{enumerate}

\item Specify $N_1$ and $N_2$, the number of pixels in the image;
$\delta x$ and $\delta y$, the size of each pixel; $C_l, l=2,\ldots,
l_{\rm max}$, a tabulated angular power spectrum; and a random number
seed.

\item Compute the cell size in the $(u,v)$ plane, $\delta u = 1/N_1
\delta x$, $\delta v = 1/N_2 \delta y$.

\item Create a complex array of size $N_1 \times N_2$, with indices
$-N_1/2 \le k_1 \le N_1/2-1$, $-N_2/2 \le k_2 \le N_2/2-1$. 

\item For each element in this array, compute $l= 2\pi\sqrt{
(k_1\delta u)^2 + (k_2\delta v)^2} -\half$, find the corresponding
$C_l$ in the tabulated power spectrum, and compute $\sigma =
\sqrt{\overline{C_l}\delta u \delta v}$ where $\overline{C_l}$ is the
sum of $1/12$ of each corner value and $8/12$ of the central value (an
approximation of the integration of $C_l$ over the cell; cf. Simpson's
rule). The $l$'s for the cell corners are given by
\begin{equation}
2\pi\sqrt{ ([(k_1\pm\half)\delta u]^2 + [(k_2\pm\half)\delta v]^2} - \half.
\end{equation}

\item Assign to both the real and imaginary parts of each element
numbers taken from a gaussian distribution $N(0, \sigma/\sqrt{2})$. To
take into account conjugate symmetry, one half of the array must be
copied from the other half, and the central $(0,0)$ element must be
real; for simplicity, we set this element (corresponding to $C_0$) to
zero so the sum of the sky image pixels is zero.

\item Perform the FFT to obtain a real (not complex) sky image of size
$N_1 \times N_2$.

\item Scale the image by $T_0 = 2.725$~K to get an image of $\Delta T$.

\end{enumerate}

\section{Calculation of X-ray density model bias correction}
\label{section:bias}

We can calculate the bias factor due to the density profiles by
showing the impact the distribution of model parameters has on the SZE
model fitting.  The best-fit model is calculated by minimizing
$\chi^2$:
\begin{equation}
\chi^2 = \sum_j \left(\frac{V_{dj}-V_{mj}}{\sigma_j}\right)^2,
\end{equation}
where $V$ represents the visibility data, which we write as
$V(u,v)\equiv I \hat{V}(u,v)$, and the index $j$ indicates a summation
over each visibility data point.  $I$ represents the overall scaling,
which is a simple function of $\theta_0$, $\beta$, and $n_{e0}$ (from
the factors that come out of the SZE volume integral and is $\propto
n_e\theta_0 \frac{\Gamma(3\beta/2 - 1/2)}{\Gamma(3\beta/2)})$);
$\hat{V}$ represents the part of the fit that depends on the shape of
the cluster and is a more complicated function of $\theta_0$ and
$\beta$ (i.e., it's the Fourier transform of the $\beta$-model image).
We use the subscripts $d$ and $m$ to represent the data and the model,
respectively.  

In the model, we define $h\equiv 1$.  $I_{d}$ has an
implicit dependence on $h^{-1/2}$, which is what we're ultimately
fitting for.  To be explicit,
\begin{equation}
I_d \propto n_d \theta_d \frac{\Gamma(3\beta_d/2 - 1/2)}{\Gamma(3\beta_d/2)} h^{-1/2}
\end{equation}
and
\begin{equation}
I_m \propto n_m \theta_m \frac{\Gamma(3\beta_m/2 - 1/2)}{\Gamma(3\beta_m/2)} (h=1).
\end{equation}
After minimizing $\chi^2$, we have
\begin{equation}
\sum\frac{I_d\hat{V}_{dj}^2}{\sigma^2_j} =
\sum\frac{I_m\hat{V}_{mj}\hat{V}_{dj}}{\sigma_j^2},
\end{equation}
which we can rewrite as
\begin{equation}
n_d \theta_d \frac{\Gamma(3\beta_d/2 - 1/2)}{\Gamma(3\beta_d/2)}
h^{-1/2}\sum\frac{\hat{V}_{dj}^2}{\sigma^2_j} = n_m \theta_m
\frac{\Gamma(3\beta_m/2 -
1/2)}{\Gamma(3\beta_m/2)}\sum\frac{\hat{V}_{mj}\hat{V}_{dj}}{\sigma_j^2}.
\end{equation}
By putting together all the above pieces, we see that the estimated
value of \hminhalf from the model-fitting is represented by
\begin{equation}
h^{-1/2} = \left <\frac{n_m \theta_m \frac{\Gamma(3\beta_m/2 -
    1/2)}{\Gamma(3\beta_m/2)}\sum\frac{\hat{V}_{mj}\hat{V}_{dj}}{\sigma_j^2}
    }{n_d \theta_d \frac{\Gamma(3\beta_d/2 -
    1/2)}{\Gamma(3\beta_d/2)}\sum\frac{\hat{V}_{dj}^2}{\sigma^2_j}}
    \right >
\end{equation}
which will not be the same as the actual value of \hminhalf if the
model parameters derived from the X-ray observations are slightly
different from those of the actual data.  To determine the bias in the
distribution of \hminhalf due to the model parameters, we calculate
the quantity ($h^{-1/2}_{obs}/h^{-1/2}_{true}$) for each of the groups
of model parameters in the distribution.  The mean of this
distribution is the X-ray model bias factor, listed in Table
\ref{tbl:errors}.

\clearpage

\begin{scriptsize}
\begin{deluxetable}{lccccccc}
\tablewidth{0pt}
\tablecaption{CBI SZE Cluster Sample\tablenotemark{1}
\label{tbl:sample}}
\tablehead{\colhead{Cluster}
&\colhead{z}
& \colhead{\begin{tabular}{c}$L_{0.1-2.4\mbox{keV}}$  \\($h^{-2}10^{44}$erg/s)$^a$\end{tabular}}
& \colhead{\hspace{0.2in}}
& \colhead{\rosat }
& \colhead{\asca}
& \colhead{\XMM}
& \colhead{\chandra}
}
\startdata
%
%
%
 A2029  &    0.773	& 3.84 & &     P	      &	  y  &	G  &S\\
 A478   &   0.0881 	& 3.24 &	&     P	      &	  y  &	G  &S\\
 A401   &   0.0737 	& 2.47 &	&     P	      &	  y  &	G  &I\\	
 A3667$^S$ & 0.0556	& 2.32 &	&     P		&  y &	G  &I\\	
 A85    &   0555 	& 2.15 &	&     P	      &	  y  &	B  &I\\	
 A3827$^S$ & 0.0984	& 1.95 & &     H              &    &	B   &\\	
 A3571  &  0.0391  	& 1.94 &	&     P	      &	  y  &	B  &\\	
 A3266$^S$ & 0.0589	& 1.89 &	&     P		&  y &	G  &I\\	
 A1651  &  0.0844   	& 1.85 &	&     P	      &	  y  &     &I   \\	
 A754   &   0.0542 	& 1.80 &	&     P	      &	  y  &	G  &I\\	
 A3112$^S$ & 0.0750	& 1.79 &	&     P		&  y &	G  &S\\	
 A399   &  0.0724  	& 1.61 &	&     P	      &	  y  &	G  &I\\	
 A1650  &  0.0845  	& 1.61 &	&     P	      &	  y  &	B  &\\
 A2597  &  0.0852  	& 1.48 &	&     P	      &	  y  &	G  &S\\
 A3558  &  0.0480  	& 1.46 &	&     P	      &	  y  &	G  &S\\
\hline
 A3695  &  0.0894  	& 1.44 &	&     H        &	     &	   &\\
PKS1550-140& 0.0970   & 1.42 &	&	&	&	&\\
  A3158$^S$& 0.0597	& 1.37 &	&     P		&  y &	 &I\\
 A3921$^S$ & 0.0936	& 1.32 &	&     P		&  y &	G &\\
  Z5029     & 0.0750& 1.32    & &	             &	     &	   &\\
  A780      & 0.0539& 1.23    &          &    P           & y & G& I,S\\
  A3911$^S$ & 0.0965& 1.23    & &     P		&    &	 &\\ 
  A2420     & 0.0846& 1.16    & &      	      &      &	   &\\
 A4010      & 0.0957& 1.16    &	&     	      &	     &	   &\\
\hline
\hline
\multicolumn{8}{l}{$^a$ XBACs and REFLEX assume $h=0.5$.  Here we convert
  their luminosities to units of $h=1.0$.}\\
\multicolumn{8}{l}{$^S$ Southern Source, not accessible with VLA or
  OVRO 40-m}\\
\multicolumn{8}{l}{\rosat: P = Public PSPC, H = Public HRI only} \\ 
\multicolumn{8}{l}{\asca: y = public data available} \\
\multicolumn{8}{l}{\XMM: G = Guaranteed Time Target, B = General Observer Target} \\
\multicolumn{8}{l}{\chandra: I = ACIS-I, S = ACIS-S} \\
\enddata
\tablenotetext{1}
{compiled from \rosat cluster surveys
  \citep{XBACS,BCS,REFLEX1,REFLEX_final}.  All redshifts are from
  \citet{SR99} except those for Z5029 \citep{BCS} and PKS1550-140
  \citep{REFLEX_final}. Luminosities are from \citet{REFLEX_final}.}
\end{deluxetable}
\end{scriptsize}


\clearpage
\begin{deluxetable}{c c c c c c c}
\tablewidth{0pc}
\tablecaption{Pointing Positions for SZE Observations\tablenotemark{1}}
\tablehead{
\colhead{Cluster} &  
\colhead{\begin{tabular}{c} R.A. \\ (J2000) \end{tabular}} &
\colhead{\begin{tabular}{c} Decl. \\ (J2000) \end{tabular}} &
\colhead{\begin{tabular}{c} \small{L\&T offsets} \\ (min) \end{tabular}} &
\colhead{\begin{tabular}{c} \small{Hours Observed} \\ (L+M+T) \end{tabular}} &
\colhead{\begin{tabular}{c} rms noise \\ \small{(mJy/beam)}\end{tabular}} &
\colhead{\begin{tabular}{c} Beam \\ FWHM \end{tabular}} 
}
\startdata
A85   &   00:41:48.7  &    $-$09:19:04.8& $\pm$16.5&16.6 &  1.8 & 5.3$'$\\
A399  &   02:57:49.7  &    +13:03:10.8& $\pm$12.5&15.6 &  2.0 &5.4$'$\\
A401  &   02:58:56.9  &    +13:34:22.8&$\pm$12.5 &15.7 &  2.0 & 5.4$'$\\
A478  &   04:13:26.2  &    +10:27:57.6&$\pm$10 &12.2 &  2.4 &5.2$'$\\
A754  &     09:09:01.4 &      -09:39:18.0&$\pm$9 &16.0 &  1.9 &5.4$'$\\ 
A1651  &  12:59:24.0   &   -04:11:20.4  &$\pm$11 &16.3 & 2.0  &4.9$'$\\
A2597  &  23:25:16.6    &  -12:07:26.4    &$\pm$15.5& 11.6 & 2.3   &5.5$'$\\ 
\enddata
\tablenotetext{1}{The offsets in
  R.A. for the Lead and Trail fields are listed in minutes.  The
  listed rms noise is for the map where the average of the Lead and
  Trail fields has been subtracted from the Main field, M-(L+T)/2.}
\label{tbl:pos}
\end{deluxetable}


\clearpage
\begin{deluxetable}{c c c c}
\tablewidth{0pc}
\tablecaption{Cluster Redshifts and Parameters derived from X-ray Observations\tablenotemark{1}}
\tablehead{
\colhead{Cluster} &  
\colhead{\begin{tabular}{c} $\theta_0$ \\ (arcm)\end{tabular}} &
\colhead{$\beta$} & 
\colhead{\begin{tabular}{c} $n_{e0}$ \\ ($10^{-3}h^{1/2}$ cm$^{-3}$)\end{tabular}}
}
\startdata
A85        &  2.04$\pm$0.52  &	  0.600$\pm$0.05   &  10.20$\pm$3.40 \\
A399       &  4.33$\pm$0.45  &	  0.742$\pm$0.042  &   3.22$\pm$0.46 \\
A401       &  2.26$\pm$0.41  &	  0.636$\pm$0.047  &   7.95$\pm$0.98 \\
A478       &  1.00$\pm$0.15  &   0.638$\pm$0.014  &  27.88$\pm$6.39 \\
A754       &  5.50$\pm$1.10  &	  0.713$\pm$0.120  &   3.79$\pm$0.07  \\
A1651      &  2.16$\pm$0.36  &	  0.712$\pm$0.036  &   6.84$\pm$1.79  \\
A2597      &  0.49$\pm$0.03  &	  0.626$\pm$0.018  &  42.99$\pm$3.82  \\
\enddata
\tablenotetext{1}{Cluster Redshifts and Parameters derived from X-ray
  Observations.  The redshifts are from the compilation of
  \citet{SR99}.  The other parameters are taken from \citet{MM2000},
  but the densities have been recalculated to account for slightly
  different temperatures, redshifts, and cosmology assumed in this
  paper.  (See text for details.)  }
\label{tbl:param}
\end{deluxetable}


\clearpage
\begin{deluxetable}{c c c c c c c c}
\tablewidth{0pc}
\tablecaption{Cluster Temperatures from \asca and \bepposax \tablenotemark{1}}
\tablehead{ 
%
& \multicolumn{2}{c}{\asca} & \asca & \bepposax & \bepposax   & \hminhalf error & Cooling Flow?}
\startdata
Cluster & White               & MFSV         & Avg   & DM2002              & \& \asca    &       & \\
        & (keV)               &  (keV)       & (keV) & (keV)               & average     &       & \\
\hline
A85  & 6.74$\pm$0.50          & 6.9$\pm$0.2         & 6.8$\pm$0.5&6.83$\pm$0.15           &{\bf 6.8$\pm$0.2} & $\pm$2.9\%        & CF\\
A399 & 6.80$\pm$0.17          & 7.0$\pm$0.2         & {\bf 6.9$\pm$0.2}&                        &            & $\pm$2.9\%        & SC\\
A401 & 8.68$\pm$0.17          & 8.0$\pm$0.2         & {\bf 8.3$\pm$0.4}&                        &            & $\pm$4.8\%        & SC\\
A478 & 7.42$^{+0.71}_{-0.54}$ & 8.4$^{+0.5}_{-0.8}$ & {\bf 7.9$\pm$0.8}&                        &            & $\pm$10.1\%        & CF\\
A754 & 9.83$\pm$0.27          & 9.5$^{+0.4}_{-0.2}$ & 9.7$\pm$0.3& 9.42$^{+0.16}_{-0.17}$ &{\bf 9.5$\pm$0.2} & $\pm$2.1\%        & SC\\
A1651& 6.21$^{+0.18}_{-0.17}$ & 6.1$\pm$0.2         & {\bf 6.2$\pm$0.2}&                        &            & $\pm$6.3\%       & SC\\
A2597& 3.91$^{+0.27}_{-0.22}$ & 4.4$^{+0.2}_{-0.4}$ & {\bf 4.2$\pm$0.4}&                        &            & $\pm$9.5\%        & CF\\
\enddata
\tablenotetext{1}{All errors are 68\% confidence. Boldfaced values are
the average electron temperatures we assumed for each cluster.  In the
final column, ``CF'' indicates that the ASCA data show a significant
central cool component in the cluster gas.  ``SC'' indicates that the
data were better fit by a single component model \citep{MFSV98}.}
\label{tbl:temps}
\end{deluxetable}


\clearpage
\begin{deluxetable}{ccccccc}
\tablewidth{0pc}
\tablecaption{Values of $h$ for A478 from isothermal fits to nonisothermal
cluster data\tablenotemark{1}}
\tablehead{
\colhead{\begin{tabular}{c}  ~~~~~~~~~~~$\gamma~~=$ \\ $r_{\rm iso}$ \end{tabular}} & 
\colhead{\begin{tabular}{c} 1.6 \\ ~\end{tabular}} &
\colhead{\begin{tabular}{c} 1.5 \\ ~\end{tabular}} &
\colhead{\begin{tabular}{c} 1.4 \\ ~\end{tabular}} &
\colhead{\begin{tabular}{c} 1.3 \\ ~\end{tabular}} &
\colhead{\begin{tabular}{c} 1.2 \\ ~\end{tabular}} &
\colhead{\begin{tabular}{c} 1.1 \\ ~\end{tabular}}
}
\startdata
0.0&  2.06&  1.81&  1.59&  1.39&  1.22&  1.10\\
0.1&  1.42&  1.33&  1.24&  1.16&  1.09&  1.04\\
0.2&  1.01&  1.00&  0.99&  0.98&  0.98&  0.98\\
0.3&  0.90&  0.91&  0.92&  0.93&  0.95&  0.97\\
0.4&  0.88&  0.89&  0.91&  0.92&  0.94&  0.97\\
0.5&  0.89&  0.90&  0.92&  0.93&  0.95&  0.97\\
0.6&  0.91&  0.92&  0.93&  0.95&  0.96&  0.98\\
0.7&  0.94&  0.94&  0.95&  0.96&  0.98&  0.99\\
0.8&  0.96&  0.96&  0.97&  0.98&  0.98&  0.99\\
0.9&  0.97&  0.98&  0.98&  0.99&  0.99&  1.00\\
1.0&  0.99&  0.99&  0.99&  0.99&  1.00&  1.00\\
\enddata
\tablenotetext{1}{$r_{\rm iso}$ is in
units of $r_{200}$.  The numbers in the DM2002 columns are boldfaced
depending on whether each cluster is a cooling flow or not.  The
boldfaced values are the ones that we use in determining the sample
error for the DM2002 mean temperature profile.}
\label{tbl:a478tprof}
\end{deluxetable}


\clearpage
\begin{deluxetable}{cccc}
\tablewidth{0pt}
\tablecaption{Nonisothermality Bias
\label{tbl:tprofsum}}
\tablehead{\colhead{\begin{tabular}{c} \\ $r_{\rm iso},\gamma$\end{tabular} }
& \colhead{\begin{tabular}{c} MFSV \\ 0,1.2\end{tabular}}
& \colhead{\begin{tabular}{c} DM (no CF)\\ 0.2,1.5\end{tabular}}
& \colhead{\begin{tabular}{c} DM (CF)\\ 0.2,1.2\end{tabular}}
}
\startdata
A85   &   1.07 & \nodata & { 1.09} \\
A399  &   1.07 & { 1.14} & \nodata  \\ 
A401  &   1.04 & { 1.12} & \nodata  \\
A478  &   0.90 & \nodata & { 1.01}  \\
A754  &   1.12 & { 1.22} & \nodata  \\
A1651 &   0.97 & { 0.99} & \nodata  \\
A2597 &   0.77 & \nodata & { 0.97}  \\
\hline
Avg \hminhalf & 0.99 & \multicolumn{2}{c}{1.08} \\
\hline
\hline
$h$ & 1.01 & \multicolumn{2}{c}{0.86} \\
\hline
\enddata

\tablenotetext{.}  { Values of $h^{-1/2}$ for the clusters in our
sample from isothermal fits to nonisothermal cluster data assuming an
input value of $h=1$.  $r_{\rm iso}$ is in units of $r_{200}$.
MFSV represents the mean temperature profile found by \citet{MFSV98},
and DM2002 are the profiles found by \citet{DM2002} for cooling flow
and non-cooling flow clusters.  The numbers in the DM2002 columns
correspond to whether each cluster is conventionally thought to
contain a cooling flow or not.}

\end{deluxetable}


\clearpage
\begin{deluxetable}{cccccc}
\tablecaption{Fit Results
\label{tbl:fit_results}}
\tablewidth{0pt}
\tablehead{\colhead{Cluster}
& \colhead{\begin{tabular}{c}Best fit $\Delta I_0$\\(mJy/arcm$^2$)\end{tabular}}
& \colhead{\begin{tabular}{c}Predicted $\Delta I_0$ \\(mJy/arcm$^2$)\end{tabular}}
& \colhead{\begin{tabular}{c}Best fit \\\hminhalf \end{tabular}}
& \colhead{\begin{tabular}{c}Reduced\\$\chi^2$\end{tabular}}
& \colhead{\begin{tabular}{c}degrees\\of freedom\end{tabular}}
}
\startdata
A85
& -1.43 
& -1.16 
& 1.24
& 1.05
& 12814
\\
A399
& -0.17 
& -0.76
& 0.23
& 1.04
& 19659
\\
A401
& -1.48
& -1.46
& 1.01
& 1.03
& 19653
\\
A478
& -4.39
& -2.49
& 1.76
& 1.06
& 20658
\\
A754
& -1.38
& -1.25
& 1.11 
& 1.06
& 15950
\\
A1651
& -1.27
& -0.89
& 1.43
& 1.14
& 19628
\\
A2597
& -1.81
& -1.05
& 1.72  
& 1.07
& 13856
\\
\enddata
\tablenotetext{.}
{The values in this table are the raw numbers obtained from the fits
  to the CBI visibility data and do not include corrections from X-ray
model bias or unsubtracted point sources.}
\end{deluxetable}

\clearpage
\begin{deluxetable}{ccccccc}
\tablecaption{Errors
\label{tbl:errors}}
\tablewidth{0pt}
\tablehead{\colhead{Cluster}
&\colhead{CMB error}
&\colhead{\begin{tabular}{c}X-ray mod\\bias \end{tabular}}
&\colhead{\begin{tabular}{c}pt src\\bias\end{tabular}}
&\colhead{\begin{tabular}{c}$T_e$\\error\end{tabular}}
&\colhead{\begin{tabular}{c}$V_{\rm pec}$\\error\end{tabular}}
&\colhead{\begin{tabular}{c}CMB+Ther+ptso\\error \end{tabular}}
}
\startdata
A85
& $\pm$0.36
& 1.01
& +0.00
& 0.03
& 0.05
& $\pm$0.38
\\
A399
& $\pm$0.42
& 1.01
& +0.02 
& 0.03
& 0.05
& $\pm$0.42
\\
A401
& $\pm$0.27
& 1.01
& +0.03
& 0.05
& 0.04
& $\pm$0.27 
\\
A478
& $\pm$0.25
& 1.00 
& +0.00
& 0.10
& 0.04
& $\pm$0.25
\\
A754
& $\pm$0.26
& 1.04
& +0.02  
& 0.02
& 0.04
& $\pm$0.29
\\
A1651
& $\pm$0.43
& 1.00
& +0.00
& 0.06
& 0.06
& $\pm$0.44
\\
A2597
& $\pm$1.06
& 1.00 
& +0.01
& 0.09
& 0.08
& $\pm$1.07
\\ \enddata \tablenotetext{.}  {The X-ray model bias is described in
Appendix \ref{section:bias}.  One divides the raw \hminhalf by this
number to correct for the bias.  The unsubtracted point source bias is
described in the text.  One adds the raw \hminhalf by this number to
correct for it.  The $T_e$ and $V_{\rm pec}$ errors are fractional;
multiply \hminhalf by these to get the error.  The CMB and the
CMB+Thermal noise+subtracted point source error listed is the absolute
error in \hminhalf.}
\end{deluxetable}

\clearpage
\begin{deluxetable}{cccc}
\tablecaption{Final Results
\label{tbl:final_results}}
\tablewidth{0pt}
\tablehead{\colhead{Cluster}
& \colhead{\begin{tabular}{c}Corrected \hminhalf\\w/ total random error\end{tabular}}
& \colhead{\begin{tabular}{c}$\Delta T_0$ \\$\mu$K\end{tabular}}
& \colhead{\begin{tabular}{c}Compton-$y_0$\\($\times 10^{-4}$)\end{tabular}}
}
\startdata
A85
& 1.23$\pm$0.40
& -580$\pm$190
& 1.13$\pm$0.37
\\
A399
& 0.24$\pm$0.42
& -80$\pm$130
& 0.15 $\pm$0.26
\\
A401
& 1.03$\pm$0.29
& -620$\pm$170
& 1.20  $\pm$0.34
\\
A478
& 1.76$\pm$0.34
& -1800$\pm$350
& 3.49$\pm$0.68
\\
A754
& 1.09$\pm$0.31
& -560$\pm$160
& 1.09$\pm$0.31
\\
A1651
& 1.42$\pm$0.47
& -520$\pm$170
& 1.00$\pm$0.33
\\
A2597
& 1.74$\pm$1.10
& -750$\pm$670
& 1.43$\pm$1.28
\\
\hline
\hline
mean $\pm$ sd =
& 1.22$\pm$0.52
&
&
\\
(probability=21\%) 
& $\frac{\chi^2}{\nu} = 1.47$ for 6 dof
&
&
\\
\hline
unweighted sample average: \hminhalf =
&1.22$\pm$0.20
&
&
\\
$\rightarrow$
&$h=0.67^{+0.30}_{-0.18}$
&
&
\\
\hline
weighted sample average: \hminhalf =
&1.16$\pm$0.14
&
&
\\
$\rightarrow$
&$h=0.75^{+0.23}_{-0.16}$
&
&

\enddata
\tablenotetext{.}
{The values in this table have been corrected for the X-ray and
  unsubtracted point source biases.  The errors listed are 68\%
  confidence random errors from the CMB anisotropies, thermal noise,
  calibration errors, point source subtraction, asphericity, temperature
  determination (within context of an isothermal model), and peculiar velocities.}
\end{deluxetable}

\begin{deluxetable}{cccc}
\tablewidth{0pt}
\tablecaption{Summary of Statistical Uncertainties in \hminhalf
\label{tbl:errsum}
}
\tablehead{\colhead{}
& \colhead{\begin{tabular}{c}Average fractional\\ error per cluster\end{tabular}}
& \colhead{\begin{tabular}{c}Average absolute \\error per cluster\end{tabular}}
& \colhead{\begin{tabular}{c}Absolute error \\for sample\end{tabular}}
}
\startdata
CMB + Therm + ptso     & 37\% & 0.45 & 0.196 \\
Asphericity            & 7\% & 0.09 & 0.035 \\
Cluster Temperature     & 6\% & 0.07 & 0.023 \\
Peculiar Velocity      & 6\% & 0.07 & 0.022 \\
Radio Calibration      & 2\% & 0.02 & 0.006 \\
\tableline		      		      
All sources            & 39\% & 0.47 & 0.202 \\
\enddata
\tablenotetext{.}  
{Fractional errors have been converted to absolute errors using our sample average of \hminhalf=1.22.  See text for details.}
\end{deluxetable}


\clearpage
\begin{deluxetable}{c c c c}
\tablewidth{0pc}
\tablecaption{Comparison of CBI $H_0$ results with \citet{MasonSZ} results\tablenotemark{1}}
\tablehead{
\colhead{Cluster} &
\colhead{\begin{tabular}{c} CBI $h^{-1/2}$ w/ \\ MMR param\end{tabular} } &
\colhead{\begin{tabular}{c} MMR $h^{-1/2}$ \\ w/ Main CMB \end{tabular}  } &
\colhead{\begin{tabular}{c} MMR \hminhalf  uncertainty \\ w/o Main CMB\end{tabular}  } 
}
\startdata
A399    & 0.23$\pm$0.26                    & 0.99$^{+0.44}_{-0.31}$ & $\pm$0.21  \\
A401    & 1.06$\pm$0.16                    & 1.40$^{+0.29}_{-0.27}$ & $\pm$0.18  \\
A478    & 1.65$\pm$0.16                    & 1.28$^{+0.28}_{-0.25}$ & $\pm$0.18  \\
A1651   & 1.47$\pm$0.27                    & 1.67$^{+0.52}_{-0.48}$ & $\pm$0.33  \\
\enddata
\tablenotetext{1}{Uncertainties listed for the CBI results only include
  statistical errors from the Lead and Trail CMB contamination, and
  errors from thermal noise in the Main field.  The MMR results list
  the quoted uncertaintiess from Table 2 of \citet{MasonSZ}.  The
  final column lists the uncertainties for the MMR results if one
  ignores the contribution to the uncertainty from the Main field
  CMB.}
\label{tbl:cf_mmr}
\end{deluxetable}


\clearpage

\end{document}